\documentclass[natbib]{svjour3}

\usepackage{mathptmx}
\usepackage{graphicx}

\usepackage{amssymb}

\journalname{Space Science Reviews}

\newcommand{\aap}{{Astron. Astrophys.}}
\newcommand{\aj}{{Astron. J.}}
\newcommand{\apj}{{Astrophys. J.}}
\newcommand{\apjl}{{Astrophys. J. Lett.}}
\newcommand{\apss}{{Astrophys. Spa. Sci.}}

\newcommand{\jgr}{{J. Geophys. Res.}}
\newcommand{\mnras}{{Monthly Not. Royal Astron. Soc.}}
\newcommand{\nat}{{Nature}}
\newcommand{\pasj}{{Publ. Astron. Soc. Japan}}
\newcommand{\solphys}{{Solar Phys.}}

\begin{document}

\title{Prominence seismology using small amplitude oscillations}

\author{Ram\'on Oliver}

\institute{Departament de F\'\i sica, Universitat de les Illes Balears, 07122 Palma de Mallorca, Spain \\ \email{ramon.oliver@uib.es}}

\date{Received: date / Accepted: date}

\maketitle

\begin{abstract}
Quiescent prominences are thin slabs of cold, dense plasma embedded in the much hotter and rarer solar corona. Although their global shape is rather irregular, they are often characterised by an internal structure consisting of a large number of thin, parallel threads piled together. Prominences often display periodic disturbances mostly observed in the Doppler displacement of spectral lines and with an amplitude typically of the order of or smaller than 2--3 km s$^{-1}$, a value which seems to be much smaller than the characteristic speeds of the prominence plasma (namely the Alfv\'en and sound velocities). Two particular features of these small amplitude prominence oscillations is that they seem to damp in a few periods and that they seem not to affect the whole prominence structure. In addition, in high spatial resolution observations, in which threads can be discerned, small amplitude oscillations appear to be clearly associated to these fine structure constituents. Prominence seismology tries to bring together the results from these observations (e.g. periods, wavelengths, damping times) and their theoretical modelling (by means of the magnetohydrodynamic theory) to gain insight into physical properties of prominences that cannot be derived from direct observation. In this paper we discuss works that have not been described in previous reviews, namely the first seismological application to solar prominences and theoretical advances on the attenuation of prominence oscillations.

\end{abstract}

\section{Introduction}

\begin{figure}
\center{\mbox{
\includegraphics[height=4.5cm]{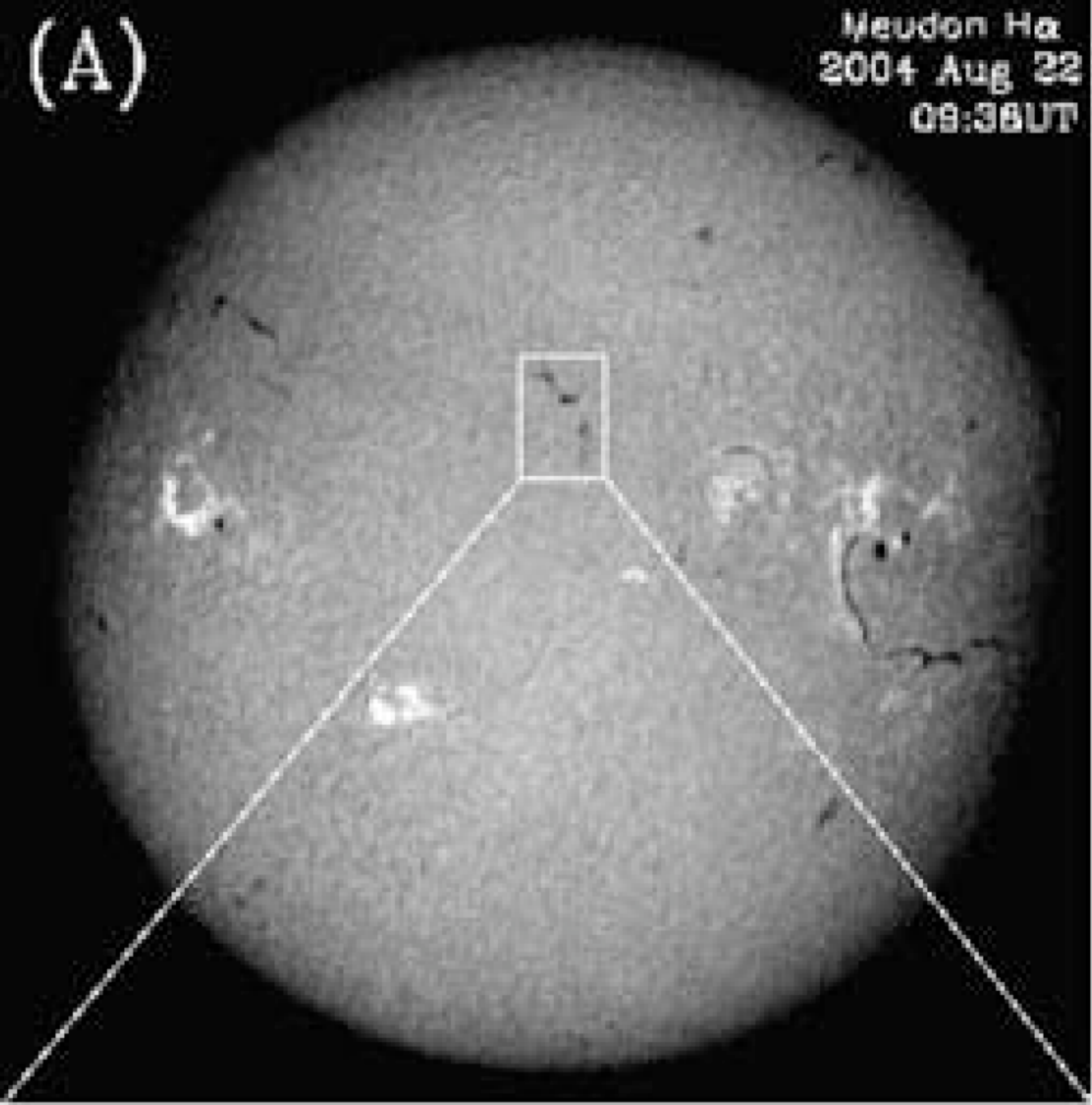}
\includegraphics[height=4.5cm]{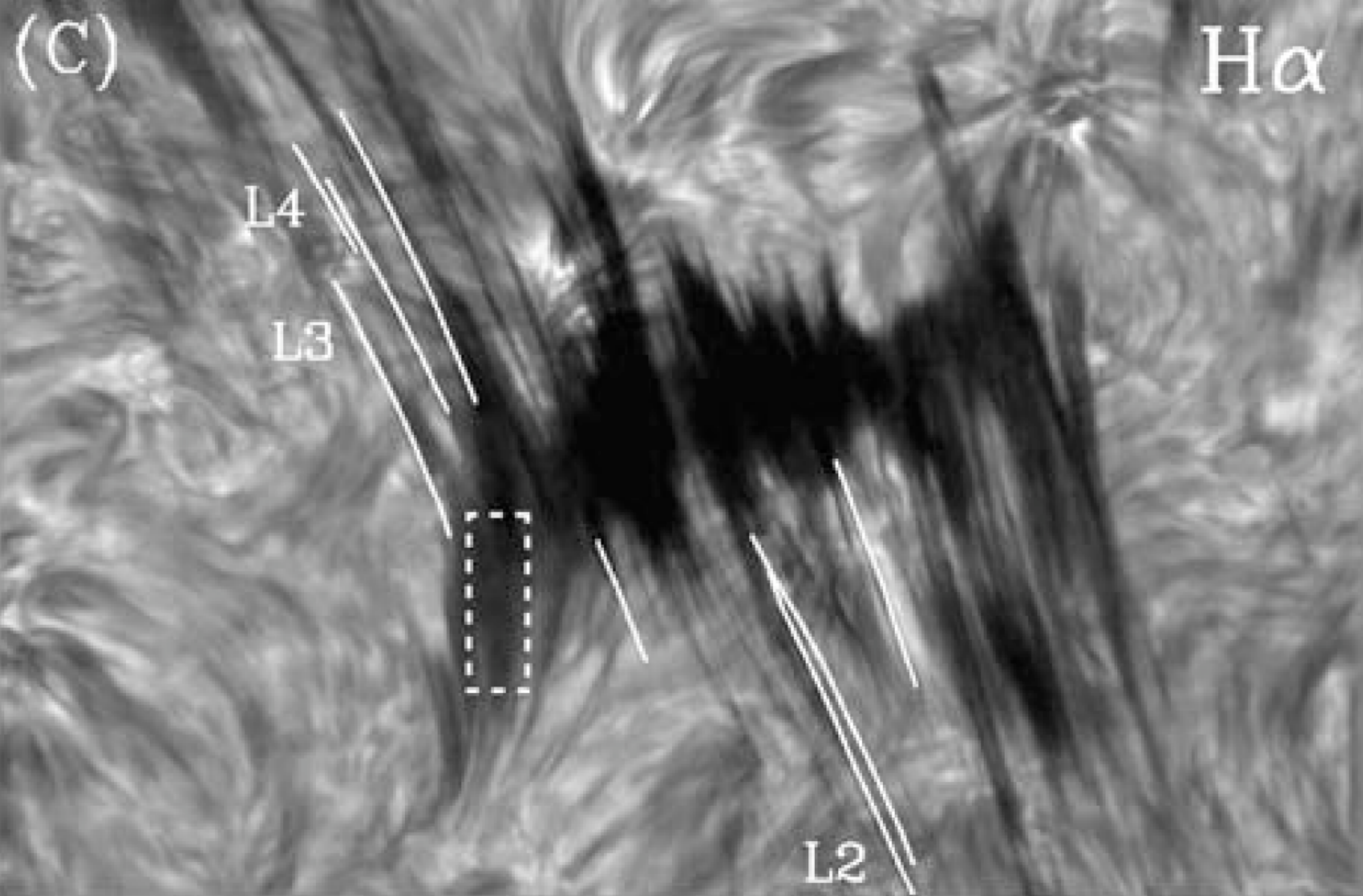}
}}
\caption{Left: H$\alpha$ image of the full solar disk showing the dark, long and thin structures known as solar prominences or filaments. Right: high-resolution image of the central part of the filament in the white rectangle of the left panel. The fine structure of the prominence, in the form of narrow threads, is clearly visible. Images taken from \cite{LEvdVvN07}.}
\label{fig1}
\end{figure}

The solar corona is often populated by peculiar objects, dense clouds of cold plasma inexplicably floating tens of thousands of kilometres above the photosphere. Such objects are routinely seen during solar eclipses, when they can be easily distinguished by their red glow, but they can also be unveiled with the help of filters, such as H$\alpha$, devised to observe the chromosphere (Fig.~\ref{fig1}). To put it in simple words, these objects (usually called prominences or filaments) are like chunks of chromospheric gas defying the downward pull of gravity and staying in a place higher than the one that apparently corresponds to their large density. This is not the only enigma around solar prominences. In contrast with the MK temperature of the surrounding corona, prominences remain at a comparatively cool 10,000~K, which prompts one to ask what prevents the mechanisms that heat the corona from also raising the temperature of prominences and consequently dispersing them. Other pieces of the prominence puzzle concern their beginning and end: first, one may wonder not just how prominences form but also why they are born in an adverse environment. Second, despite their internal dynamics, prominences that have been stable and healthy for weeks suddenly die in an spectacular eruption. Hence, like all living beings these creatures are born, grow, (do not reproduce) and die, but unlike many living beings the processes shaping the lifetime of prominences are largely unknown. Nevertheless, the intervention of a decisive element is quite clear: the magnetic field, that is central to all the mentioned processes.

The reason of our limited insight into the nature of prominences probably has three causes \citep{V98}: there is no such thing as a canonical prominence, but a wide range of parameters is observed in different objects; no prominence has a uniform structure, but they are made of thin threads (or fibrils) and, in addition, different parameter values can be detected in different parts of a prominence; and no structure is really isolated, so it is necessary to understand the physics of the prominence-corona interface, the effect of the coronal radiation field \citep[e.g.][]{AH05} and to trace the magnetic fields permeating the prominence to their origin at the Sun's surface \cite[e.g.][]{LWEvdVF05b}. Our knowledge about prominences has been well reviewed by \cite{TH95,M98,PV02}, where ample information on the topic can be found.

One of the first studies about prominence oscillations was carried out by \cite{H69}, who noted that in a sample of 68 non-active region prominences, 31\% of the objects presented no significant velocity change along the line-of-sight, 28\% showed apparently random line-of-sight velocity variations and 41\% presented a definite oscillatory behaviour. Similar results were obtained for a set of 45 active region prominences. There are several reasons that may lead to the absence of periodic variations in some prominences, e.g. the velocity amplitude or its projection along the line-of-sight are too small to stand above the instrumental noise level; or the prominence material does not actually oscillate at the time the observations are performed; or the light emitted or absorbed by various plasma elements along the line-of-sight and having different oscillatory properties results in a noisy signal. After this leading work the subject of prominence oscillations remained dormant for about 15 years and revived in the 1980s, when a great amount of observations about prominence oscillations started being collected.

According to the oscillatory amplitude, these events have been classified into two groups \citep{OB02}:

\begin{itemize}

\item {\em Large amplitude oscillations}, that arise when the whole prominence is shaken by a Moreton wave \citep{M60} impinging on it \citep{RS66}. As a consequence, the prominence gas undergoes a large displacement from its equilibrium position and the prominence as a whole vibrates with considerable velocity amplitude (of the order of 20~km~s$^{-1}$ or higher). These phenomena are reviewed in a separate work of this volume (Tripathi et al. 2009).

\item {\em Small amplitude oscillations}, with velocity amplitudes much smaller than those of large amplitude oscillations have also been frequently observed. The detected peak velocity ranges from the noise level (down to 0.1 km~s$^{-1}$ in some cases) to 2--3~km~s$^{-1}$, although larger values have also been reported \citep{BM84, MWBOB99}.

\end{itemize}

A distinguishing property of small amplitude oscillations, which gives them a distinction from large amplitude ones, is that they seldom influence the entire prominence. Thus, when spectrographic observations are carried out with a slit it is usually found that only a few consecutive points along the slit present time variations with a definite period, while all other points lack any kind of periodicity \cite[e.g.][]{TT86,BW94,MOBB97}. \cite{TMWBOB02} made a comprehensive study of the two-dimensional spatial distribution of Doppler velocity oscillations in a very large polar crown prominence and showed that, in their particular observation, oscillations are concentrated in a restricted area (54,000 $\times$ 40,000 km in size). Both propagating waves and standing oscillations are detected in this region and their wavelength and phase speed are determined. Using Hinode SOT observations of a quiescent prominence, \cite{berger_etal08} also report on oscillations that do not affect the whole prominence body. These oscillations have periods from 20 to 40 minutes and typically last for only one or two cycles. They correspond to waves propagating vertically along threads with a projected phase speed of $\sim10$ km s$^{-1}$.

It has been mentioned that solar prominences are constituted by many thin, parallel magnetic threads filled with cold plasma, and as a consequence the dynamics of these components can form the basis of the dynamics of small amplitude oscillations. Early works \citep{YEK91,YE91} already noted the possible link between small amplitude prominence oscillations and the fibril structure. Unfortunately, the spatial resolution of the data analysed by \cite{TMWBOB02} is not good enough to distinguish the prominence threads. It was necessary, hence, to wait until the advent of telescopes with much better spatial resolution to have observations in which the prominence fine structure is well resolved \citep{LEW03,L04}. In the analysis of the Doppler velocity in two threads belonging to the same filament, \cite{L04} finds a clear sign of propagating waves and determines their period, wavelength and phase speed. This study is followed by a much more profound one in which the two-dimensional motions and Doppler shifts of 328 features (or ``blobs'') of different threads are examined. These features are observed to flow along the filament axis while oscillating at the same time. To simplify the examination of oscillations, \cite{L04} computed average Doppler signals for each fibril and found that groups of adjacent threads oscillate in phase with the same period. This has two consequences: first, since the periodicity is outstanding in the averaged signal for each thread, the wavelength of oscillations is larger than the length of the thread. Second, fibrils have a tendency to vibrate bodily, in groups, rather than independently. H$\alpha$ observations conducted with the Swedish 1-m Solar Telescope by \cite{LEvdVvN07} lead to similar results concerning the collective dynamics of fibrils, although propagating Doppler velocity signals with various periods and wavelengths in other threads of the same filament are also detected. All these observations seem to indicate that prominence fibrils sometimes support collective oscillations and sometimes oscillate on their own. This topic deserves a more detailed observational study and, given the simplicity of fibrils compared to the full filament structure, theoretical investigations can give rise to fruitful comparisons with observations.

During the 1980s some observational works, e.g. those by \cite{TT86,WBS89}, provided evidence about a temporal decrease of the oscillatory amplitude, which suggests that prominence oscillations are attenuated in time. A very clear example of this phenomenon can also be found in Figure 5a of \cite{LEL77}. Here the integrated line intensity of the He D$_3$ line displays oscillations with a period around 25 min and with an amplitude that decreases in time. These oscillations are present for a few cycles but, unfortunately, a precise value of the attenuation rate is not computed in this case \citep[Figure~3 of ][also presents similar results, although with a much better spatial and temporal resolution]{berger_etal08}. More detailed observational analyses of damped prominence oscillations can be found in \cite{MWBOB99,TMWBOB02}, who studied the spatial and temporal features of Doppler velocity oscillations in limb prominences. In the second of these works, already mentioned before, a sinusoidal function multiplied by a factor $\exp(-t/\tau)$ is fitted to the Doppler series, which allowed these authors to obtain values of the damping time, $\tau$, which are usually between 1 and 3 times the corresponding oscillatory period. In spite of the lack of similar studies, the existing evidence suggests that small-amplitude oscillations in prominences are excited locally and are damped in a few periods by an unknown mechanism.

Many previous observational and theoretical results on prominence oscillations have been reviewed by e.g. \cite{OB02,E04,W04,B06,BEOO07,E08}, so our purpose here is to extend these review works by summarising other papers not discussed by these authors. Thus, we describe the first seismological analysis of prominence oscillations (Section~\ref{seismology}) and the theoretical studies on the attenuation of small amplitude oscillations (Sections~\ref{leakage}, \ref{nonideal} and \ref{resabs}). The plausible theoretical explanations of this effect are reviewed here and the relevance of different mechanisms is assessed on the basis of the ratio of the damping time to the period ($\tau/P$), that must be in the range 1--10 to agree with the observational results. Works that invoke wave leakage from the prominence as the cause for the observed damping are described in Section~\ref{leakage}, attenuation caused by non-ideal effects is discussed in Section~\ref{nonideal} and wave damping by resonant absorption is presented in Section~\ref{resabs}. Finally, Section~\ref{conclusions} contains the concluding remarks.

\section{Seismology of active region threads observed with Hinode}\label{seismology}

Observations with the Hinode Solar Optical Telescope by \cite{okamoto_etal07} display an active region prominence composed of many fine horizontal threads that oscillate independently in the vertical direction as they move along a path parallel to the photosphere. Six different threads are studied and in each of them several points are considered. All points in each thread are found to oscillate in phase. The relevant parameters for a seismological analysis of these events are flow velocities in the range 15--46 km s$^{-1}$, oscillatory periods in the range 135--250 s and thread lengths in the range 1700--16,000 km. A prominence thread is a cold plasma condensation that occupies a segment of a much longer magnetic tube. In the present case it is not easy to directly measure the length of such magnetic tubes, although \cite{okamoto_etal07} estimate that the wavelength of the oscillations is at least 250,000 km and so the minimum length of the magnetic tubes is 125,000 km. \cite{TAOB08} provide another estimation of this quantity, which they argue must be at least 100,000 km.

These authors perform an independent seismological study of the six events by assuming that the thread is a dense plasma moving along a horizontal and straight magnetic tube which is tied to the dense photosphere at its ends. Additionally the low-$\beta$ and linear approximations are also imposed. For each thread three different studies are done; in increasing order of complexity they are: (a) thin tube approximation and no flow; (b) thin tube approximation and flow; and (c) full ideal MHD equations with flow. While the thin tube approximation is well supported by the thinness of the threads compared to their length, ignoring the effect of the flow does not seem a good idea because the detected flow velocities are rather large. Nevertheless, as we describe below the results prove just the contrary.

\begin{figure}
\center{\mbox{
\includegraphics[height=4cm]{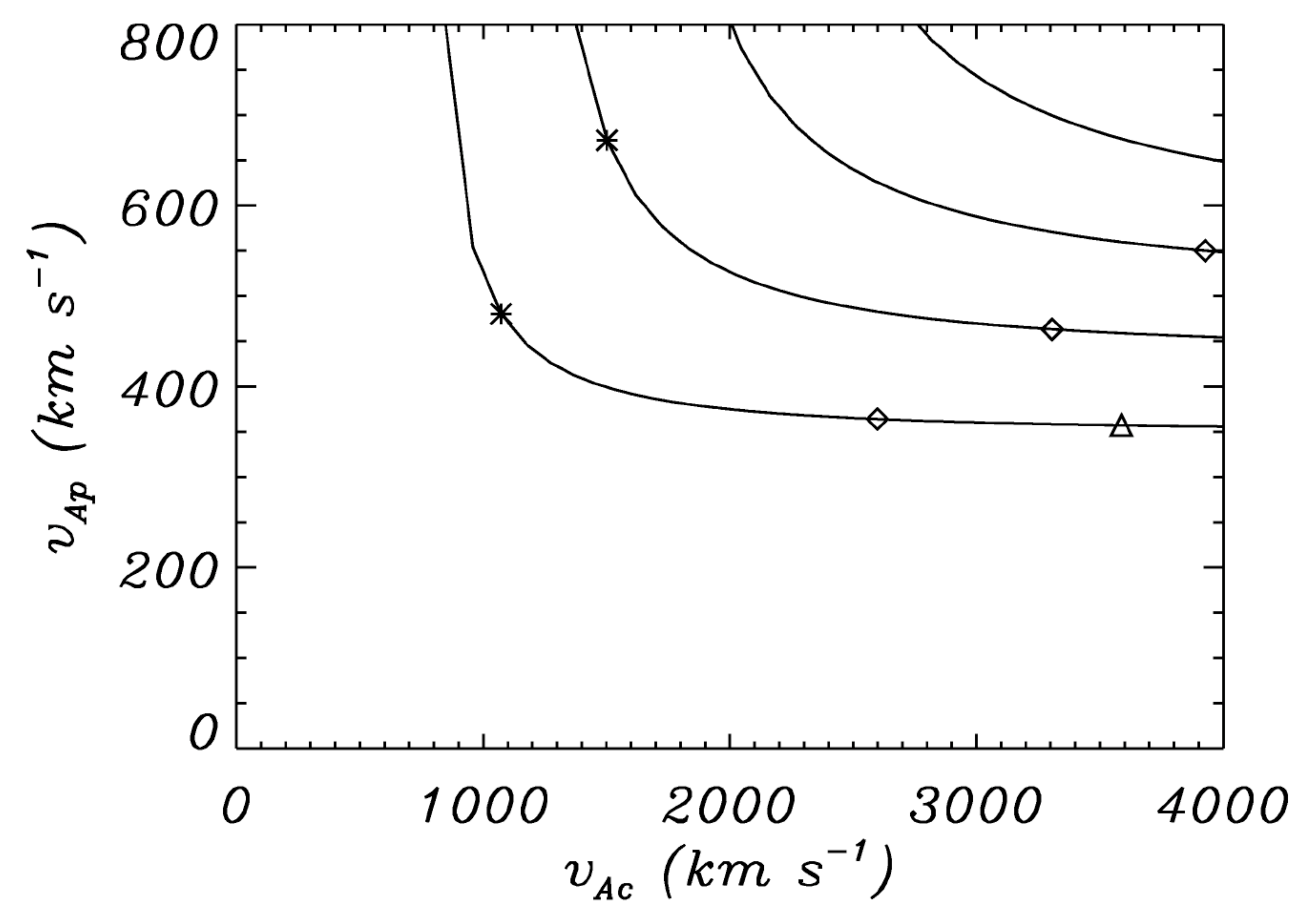}
\includegraphics[height=4cm]{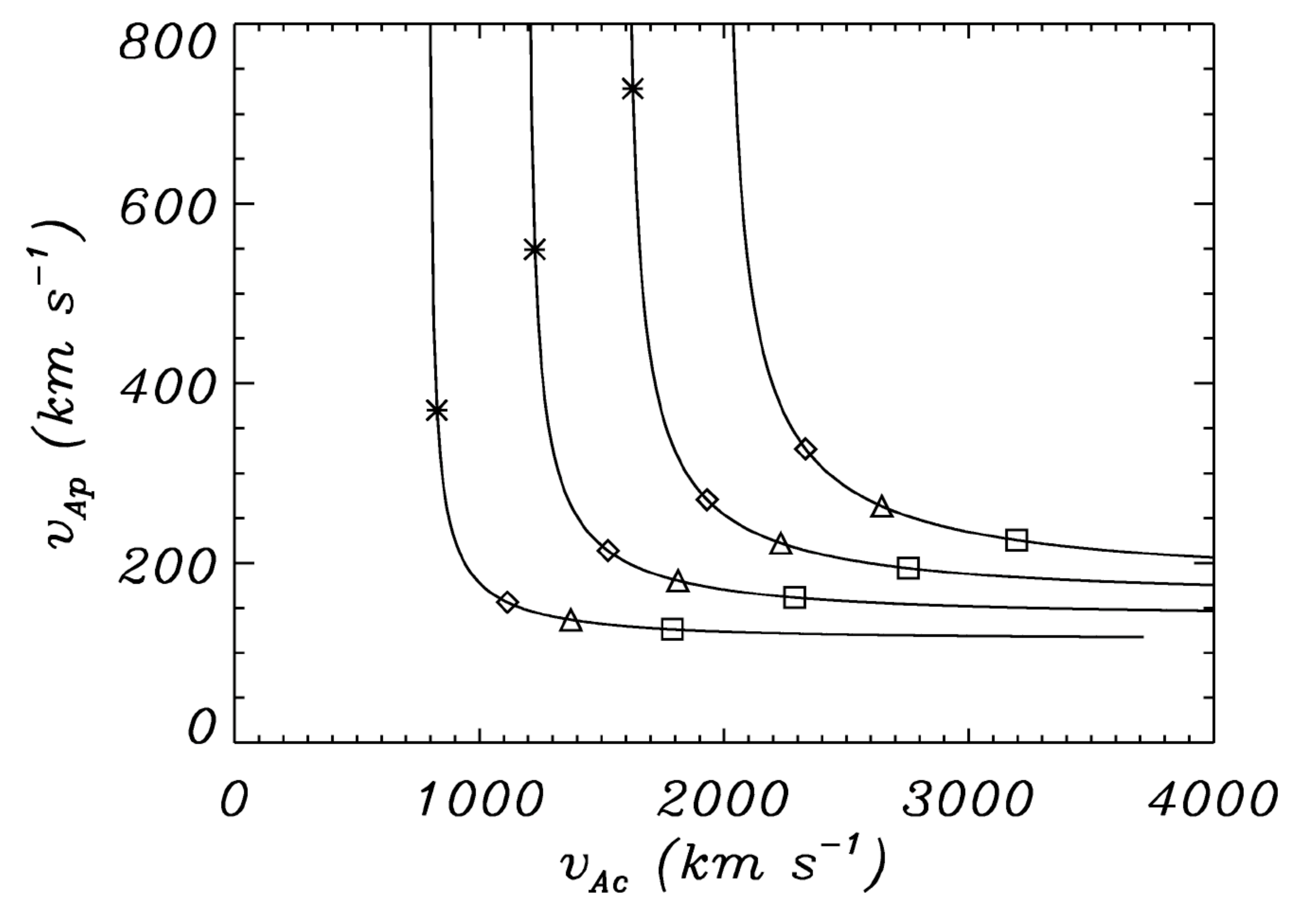}
}}
\caption{Seismology of active region threads observed with Hinode. Thread ($v_{Ap}$) versus coronal ($v_{Ac}$) Alfv\'en speed in two of the threads studied by \cite{okamoto_etal07}. The magnetic tube length ranges from 100,000 km (bottom curve) to 250,000 km (top curve) in steps of 50,000 km. Along each curve the density ratio of thread to coronal plasma is 5 ($*$), 50 ($\Diamond$), 100 ($\triangle$), 200 ($\Box$). Figures taken from \cite{TAOB08}.}
\label{dr_hinode}
\end{figure}

In the absence of flow, case (a), transverse motions of the kind observed by \cite{okamoto_etal07} can only be produced by the kink mode \citep{DR05}. The dispersion relation of the kink mode in a magnetic tube containing prominence material can be deduced from Equation~(27) of \cite{DR05}. \cite{TAOB08} conclude that, if no assumptions about the magnetic field, density and magnetic tube length ($L$) are made, then this dispersion relation contains too many unknowns and cannot be uniquely solved. But if a value of $L$ is assumed (with the restriction $L\gtrsim 100,000$ km from observations), then a dependence of the prominence versus the coronal Alfv\'en speed can be obtained. Figure~\ref{dr_hinode} displays such a dependence for two of the six threads and several magnetic tube lengths. This figure shows that, unless the ratio of prominence to coronal density is unrealistically low (e.g. below 50), the Alfv\'en speed in the thread reaches a minimum value that is not modified by that density ratio. This lower limit of $v_{Ap}$ is around 120 km s$^{-1}$ for thread \#6 and lies between 200 and 350 km s$^{-1}$ for the other five threads.

To test the validity of this seismological result, \cite{TAOB08} consider case (b), which involves solving a simple partial differential equation. Their numerical simulations of a thread that moves bodily along the magnetic tube yield in-phase oscillations of the whole thread, so a kink-like behaviour is recovered even in the presence of flow. These results are in excellent agreement with the observations of \cite{okamoto_etal07}. Perhaps surprisingly, for all six threads the period is only slightly smaller than that of case (a), the difference being below 5\%, i.e. smaller than the error bars of observations (of the order of or larger than 6.8\%). Hence, the period of these transverse oscillations is almost insensitive to the thread flowing motions. Finally, \cite{TAOB08} also investigate case (c) and recover the results of case (b), which implies that the thin tube approximation is quite reasonable, as expected. After these two studies the lower bounds for the Alfv\'en speed derived from case (a) are firmly established and although this investigation does not lead to particular values of physical parameters, such Alfv\'en speeds agree with the intense magnetic fields and large densities usually found in active region prominences.



\section{Damping of prominence oscillations by wave leakage}\label{leakage}

Wave leakage is a common phenomenon by which the energy imparted by a disturbance to a given structure (e.g. a coronal loop or a solar prominence) is not confined as standing oscillations of the structure, which emits waves carrying energy into the external medium. This situation has been studied at length in relation with coronal loop oscillations; see \cite{C86,C03} and references therein. Most of the works described in this section make no mention of wave leakage, but our interpretation is that the obtained attenuation of oscillations is simply a consequence of the emission of waves from the prominence into the corona.

\subsection{Line-current filament models}

In the celebrated model of \cite{KR74} a prominence is treated as a horizontal line current suspended in the corona. Such an approximation is justified by the fact that if the prominence plasma is supported against gravity by the Lorentz force, then a current runs along the filament. In this work the photosphere is substituted by a rigid, perfectly conducting plane with a surface current distribution caused by the coronal magnetic field generated by the filament current. Such induced currents in turn give rise to a coronal magnetic field whose effect on the prominence is to provide the lifting force necessary to counteract gravity. The full magnetic field arrangement in the corona is determined by both the line current and the photospheric currents and can be of two types: either of normal polarity (NP) or of inverse polarity (IP); see \cite{S97b,vdOSK98} for more details.

\begin{figure}
\center{\mbox{
\includegraphics[height=3.8cm]{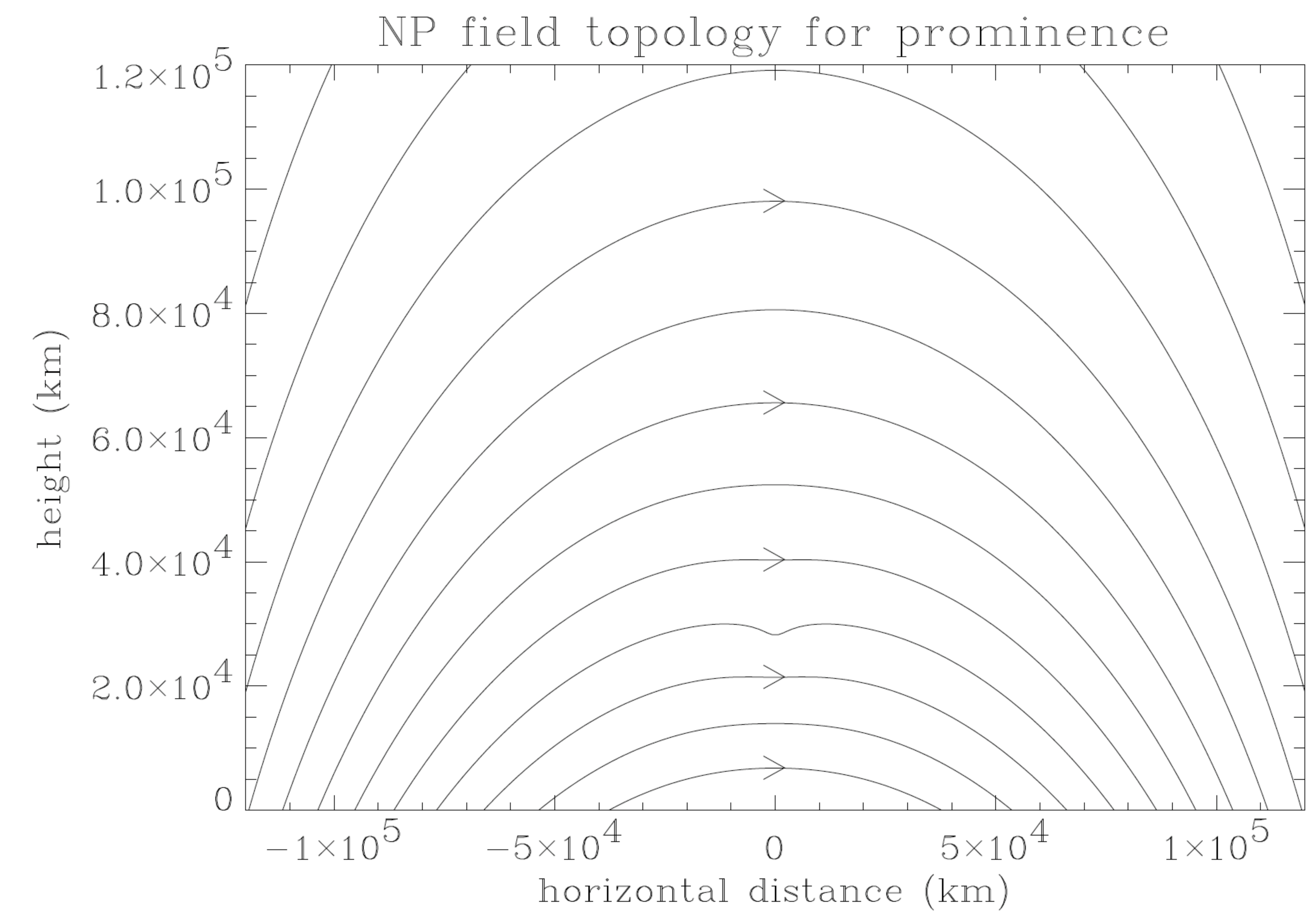}
\includegraphics[height=3.8cm]{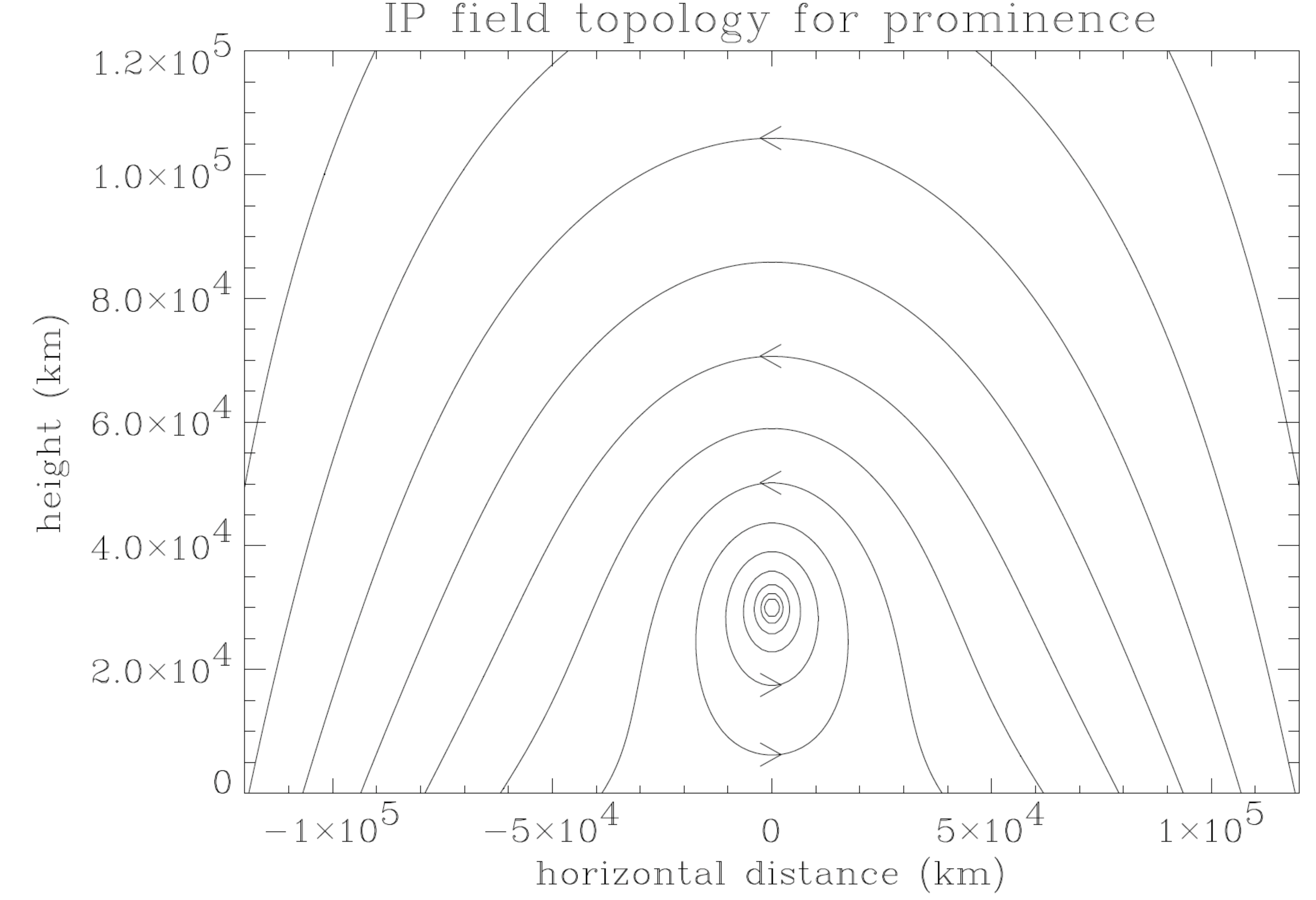}
}}
\caption{Magnetic field configuration of (a) normal polarity and (b) inverse polarity line-current prominence models used to investigate the damping of oscillations by wave leakage. The prominence electric current lies on the dip of magnetic field lines at the centre of the arcade in (a) and at the center of the magnetic island in (b). The magnetic field component perpendicular to the plane of the figure is zero. Figures taken from \cite{S97b}.}
\label{fig_np_ip}
\end{figure}

This magnetic configuration was used by \cite{vdOK92,S97a,S97b,vdOSK98} to study the oscillations of a filament. It is worth to remark that in these works the prominence is treated as an infinitely thin and long line, so that it has no internal structure, although the interplay of the filament current with the surrounding magnetic arcade and photosphere is taken into account. In addition, both NP and IP structures were considered; see Figure~\ref{fig_np_ip}, in which the poloidal magnetic field is displayed (the toroidal magnetic field component being zero). The fundamental ingredient of these papers is that if a disturbance causes a displacement of the whole line current, that remains parallel to the photosphere during its motion, then the coronal magnetic field is reshaped and the photospheric surface current is modified. This, in turn, influences the magnetic force acting on the filament current. Such a force can either strengthen the initial perturbation, so that the original equilibrium is unstable, or diminish it, so that the system is stable against the initial disturbance. This simple description becomes more complicated when one takes into account that perturbations travel at a finite speed (namely the Alfv\'en velocity) which causes time delays in the communication between the line current and the photosphere. \cite{vdOK92,S97a,S97b,vdOSK98} investigated the effect of these time delays on the filament dynamics.

\begin{figure}
\center{\mbox{
\includegraphics[height=3.8cm]{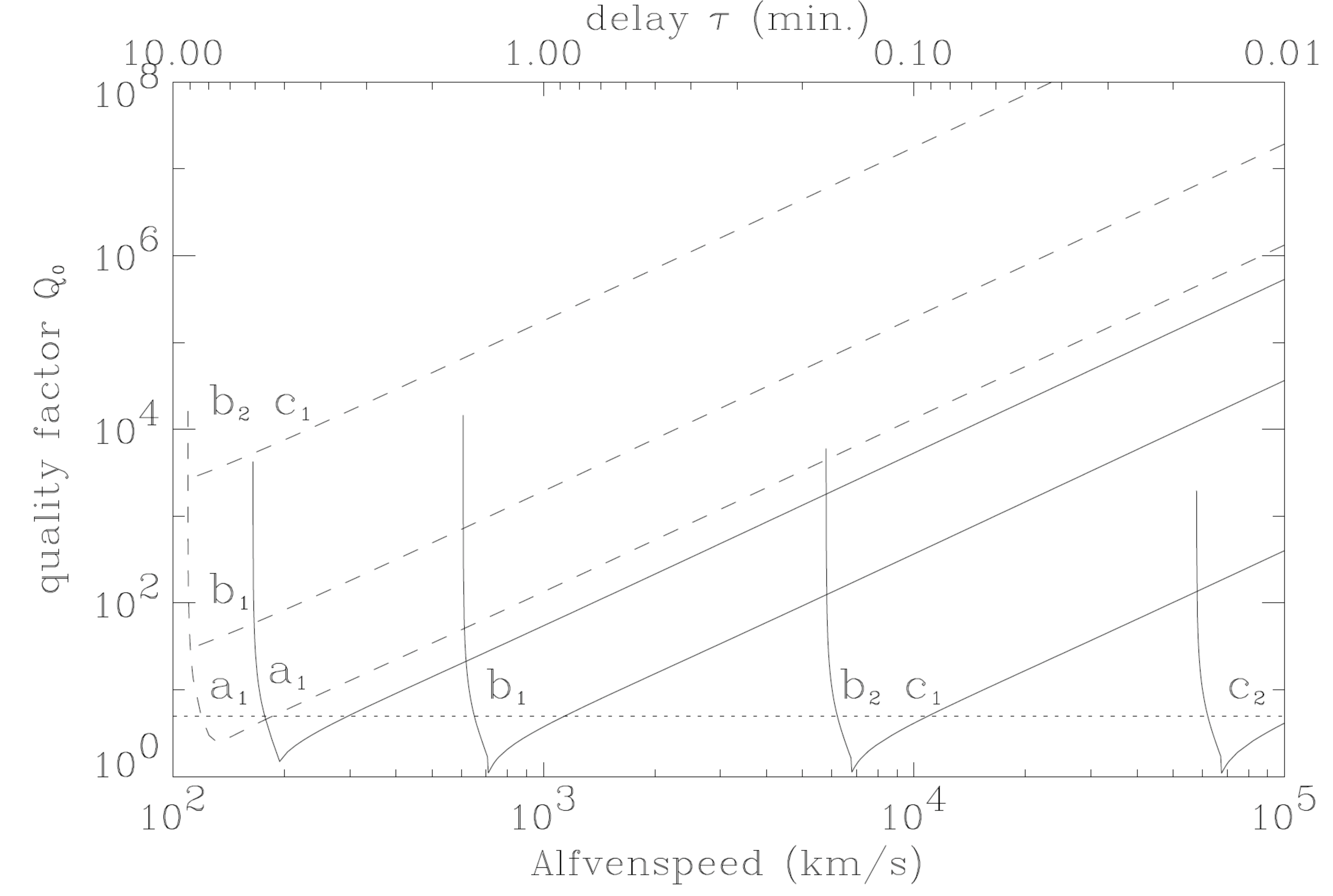}
\includegraphics[height=3.8cm]{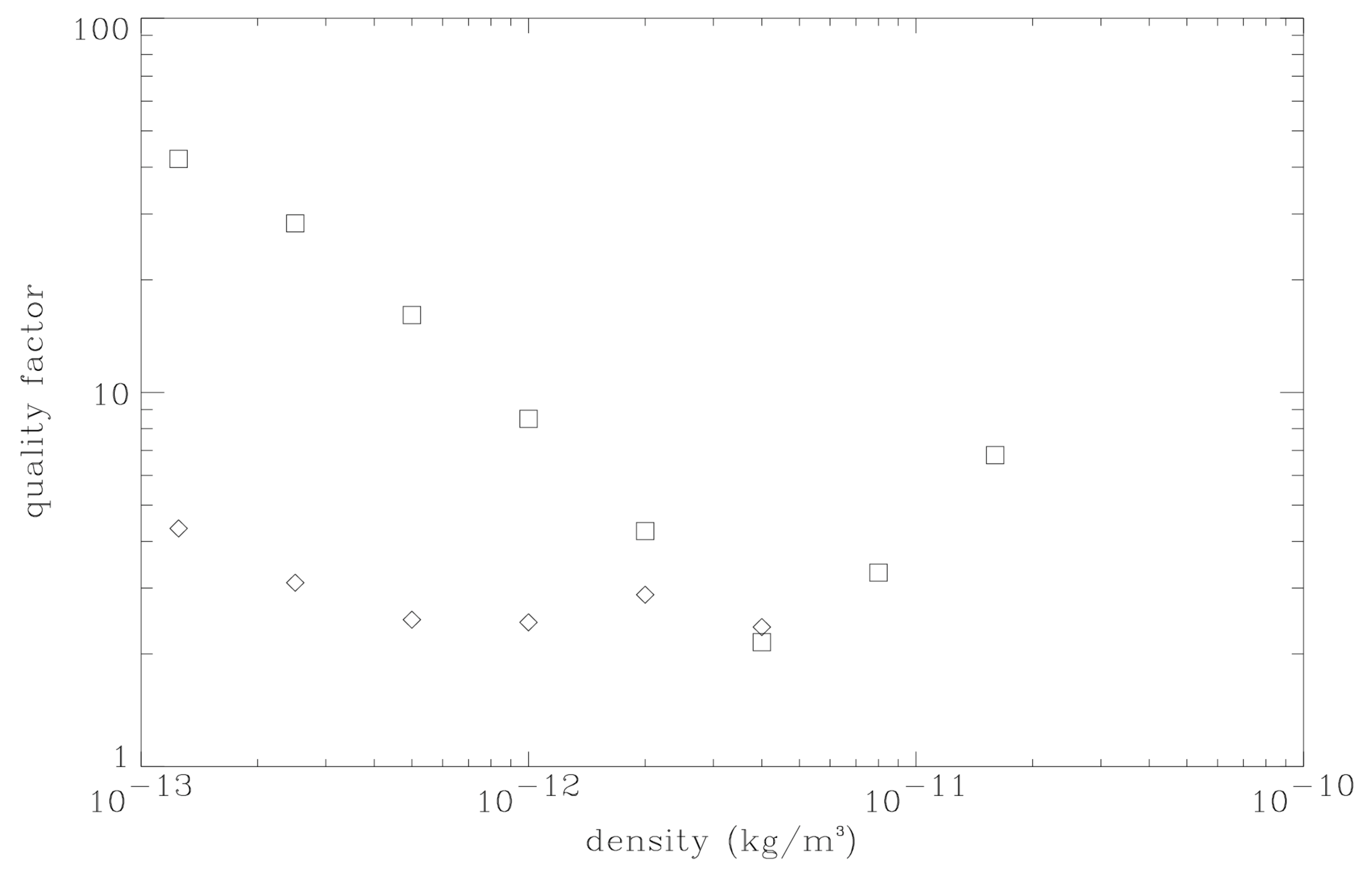}
}}
\caption{Attenuation of prominence oscillations by wave leakage. (a) Quality factor ($Q_0\equiv\pi\tau/P$) of stable IP (solid curves) and NP (dashed curves) prominence oscillations as a function of the coronal Alfv\'en speed. (b) Quality factor of the horizontally (squares) and vertically (diamonds) polarised stable oscillations versus the coronal density. Figures taken from \cite{S97b,ST99}.}
\label{fig_leakage}
\end{figure}

Exponentially growing or decaying solutions were found both for NP and IP prominence models; see Figure~4 of \cite{vdOK92} and Figure~5 of \cite{S97a} as examples. An important conclusion that can be extracted here is that the attenuation is very efficient for many parameter values, and consequently the ratio of the damping time to the period is between 1 and 10 (i.e. in agreement with observations). This point is well illustrated by Figure~\ref{fig_leakage}a, in which the quality factor is depicted as a function of the Alfv\'en speed. Different curves in this diagram correspond to different sets of parameters, with solid and dashed lines used to distinguish between IP and NP configurations. All the IP curves and one of the NP curves in this figure take values below $Q_0=30$, equivalent to $\tau/P\simeq 10$, which is an indication of the efficiency of the damping of oscillations.

\subsection{Finite-thickness filament model}

An IP magnetic configuration similar to that used by \cite{vdOK92,S97a,S97b,vdOSK98} was taken by \cite{ST99}, although in this study the prominence is not infinitely thin and instead is represented by a current-carrying cylinder. \cite{ST99} carried out numerical simulations of the isothermal magnetohydrodynamic (MHD) equations and this implies that the equilibrium temperature is set to a constant value ($10^6$ K) everywhere. The authors mention that, despite the large temperature difference between solar prominences and the corona, this isothermal assumption can only have a minor effect. In addition, the photosphere is described as a perfectly conducting boundary, as in the papers examined above. The inner part of the filament is disturbed by a perturbation with a velocity amplitude of 10 km s$^{-1}$ and at an angle of $45^{\rm o}$ to the photosphere. This causes the prominence to move like a rigid body in the corona, both vertically and horizontally, and to undergo exponentially damped oscillations; see Figure~7 in \cite{ST99}. It is found that the horizontal and vertical motions are decoupled from one another (and so can be investigated separately) and that the period and damping time of horizontal oscillations are much larger than their respective counterparts for vertical oscillations. Again, strong damping can be achieved for some parameter values; see Figure~\ref{fig_leakage}b. It turns out that vertical oscillations are very efficiently attenuated for all the parameters considered in this work and that the same happens with horizontal oscillations for coronal densities above $\simeq 5\times10^{-13}$ kg m$^{-3}$. These contrasting properties of damped horizontal and vertical oscillations contain some potential for performing seismology of prominences.

Now we turn our attention to the interpretation of the results presented so far. None of the works cited in the first part of this section, i.e. \cite{vdOK92,S97a,S97b,vdOSK98}, mention wave leakage as the cause of the attenuation, but instead these papers connect this behaviour with the effect of the distant photosphere and the presence of time delays in the communication of disturbances between the filament and the photosphere. On the other hand, \cite{ST99} link the damping of oscillations to the emission of waves by the prominence: the damping of horizontal motions is attributed to the emission of slow waves, whereas fast waves are invoked as the cause of the damping of vertical motions. Given that the main difference between this work and the other ones lies essentially in the cross section of the filament, there is no reason to believe that the physics involved are much different regardless of the prominence being modeled either as a straight and infinitely thin current or as a current-carrying cylinder. This issue must be examined in more detail to better understand the dynamics of solar prominences.

A final remark about the exact nature of the wave leakage found by \cite{ST99} must be made. In this work (see their Figure~6a) the plasma-$\beta$ in the prominence ranges from $\beta>1$ in its central part to $\beta<0.1$ at its boundary. Hence, waves emitted by the prominence into the corona propagate in a $\beta\ll 1$ environment in which magnetic field lines are closed. Under these conditions, slow modes propagate along magnetic field lines and are unable to transfer energy from the prominence into the corona and so wave leakage in the system studied by \cite{ST99} is only possible by fast waves. Then, it is hard to understand how the prominence oscillations can be damped by the emission of slow waves in this particular model, in which the dense, cool plasma is only allowed to emit fast waves. It must be mentioned, however, that the plasma-$\beta$ in the corona increases with the distance from the filament, which implies that the emitted fast waves can transform into slow waves when they traverse the $\beta\simeq1$ region. This effect has been explored by \cite{MH06,MH07}, but see also references therein for similar studies.

\section{Damping of small amplitude prominence oscillations by non-ideal effects}\label{nonideal}

\cite{B03} took into account some dissipative mechanisms and explored, through some order-of-magnitude calculations, their importance as possible damping agents. In this work we are warned about the simplicity of the calculations and about the neglect of the solar corona, whose role on the damping of prominence oscillations is beyond this simple study. Dissipative mechanisms are separated into isotropic and anisotropic. Amongst the first ones, viscosity and magnetic diffusion are found to be rather inefficient in attenuating perturbations in only a few periods, as demanded by observations. Moreover, radiative losses (here modeled by means of Newton's cooling law, that gives an extremely simplified account of radiation in a solar prominence), may be of some importance, although their effect is difficult to establish unambiguously because of the presence of the unknown radiative relaxation time-scale in Newton's cooling law. Therefore, radiation by the prominence plasma requires a more specific treatment to determine its significance. Regarding anisotropic mechanisms, in \cite{B03} it is determined that, because of the very large density and very low temperature of prominences, viscosity can be considered isotropic and thus irrelevant as the cause of the observed damping of oscillations. In addition, thermal conduction is dominated by the presence of a magnetic field and is essentially parallel to the field direction. This mechanism is very efficient as a damping agent for very short wavelengths. Since Ballai's work has been extended in other papers discussed below, the meaning of ``very short wavelengths'' will become clear later. Although other damping mechanisms are not contemplated in \cite{B03}, e.g. wave leakage, ion-neutral collisions or resonant absorption, this work is important since it allowed subsequent investigations to discard the irrelevant mechanisms and to concentrate only in the rest.

\subsection{Radiative losses and thermal conduction in a uniform medium}\label{non-adiab-unif-medium}

The efficiency of thermal conduction (parallel to the magnetic field) and radiation in transporting heat, and thus in causing the attenuation of oscillations, has been addressed in a number of papers; see \cite{IE93,COB04,TCOB05,CTOB06}. These works are concerned with the spatial and temporal damping of fast and slow waves propagating in a uniform medium. Radiative losses are simulated by Equation~(7) of \cite{H74}, which contains two parameters that can be tuned to represent optically thin or thick radiation; see \cite{COB04}. Different sources of plasma heating are also included, although this mechanism has a negligible consequence on the properties of the waves.

To understand the results of these studies it is worth to consider the characteristic time-scales of thermal conduction and radiative losses ($\tau_c$ and $\tau_r$), that following e.g. \cite{DH04} can be defined as

\begin{equation}\label{tau_c}
\tau_c = \frac{L p}{(\gamma-1)\kappa_\| T},
\end{equation}

\begin{equation}\label{tau_r}
\tau_r = \frac{\gamma p}{(\gamma-1)\rho^2\chi T^\alpha}.
\end{equation}

\noindent In these expressions $p$, $\rho$ and $T$ are the plasma pressure, density and temperature, $L$ is a characteristic length-scale (the wavelength of oscillations, for example), $\kappa_\|$ the parallel thermal conductivity and $\chi$ and $\alpha$ are the two parameters in Hildner's radiative loss function. We see that $\tau_r$ is independent of $L$ and so the rate at which the plasma looses energy through radiation does not vary with the wavelength or frequency of the waves. On the other hand, the conduction time-scale, $\tau_c$, decreases linearly with $L$ and so conduction becomes more effective at transporting heat for short wavelengths, just as noticed by \cite{B03}, for example. In the limit of very short wavelengths, conduction is so efficient that the isothermal regime is reached. The critical wavenumber, $k_c$, at which this transition occurs is given by \cite{PKS94} as

\begin{equation}
k_{c-slow} = \frac{2n k_B c_s}{\kappa_\|\cos\theta}, \hspace{1cm} k_{c-fast} = \frac{2n k_B v_A}{\kappa_\|\cos^2\theta},
\end{equation}

\noindent for slow and fast waves, respectively. Here $n$ is the number density, $k_B$ Boltzmann's constant, $c_s$ and $v_A$ the sound and Alfv\'en speeds and $\theta$ the angle between the wavenumber and the unperturbed magnetic field. To derive these formulas it is necessary to assume $c_s\ll v_A$, together with $\omega_R\approx k_xc_s$ for the slow mode and $\omega_R \approx kv_A$ for the fast mode, where $\omega_R$ is the real part of the frequency and $k_x$ the component of the wavenumber parallel to the equilibrium magnetic field.

\begin{figure}\sidecaption
\resizebox{0.5\hsize}{!}{\includegraphics*{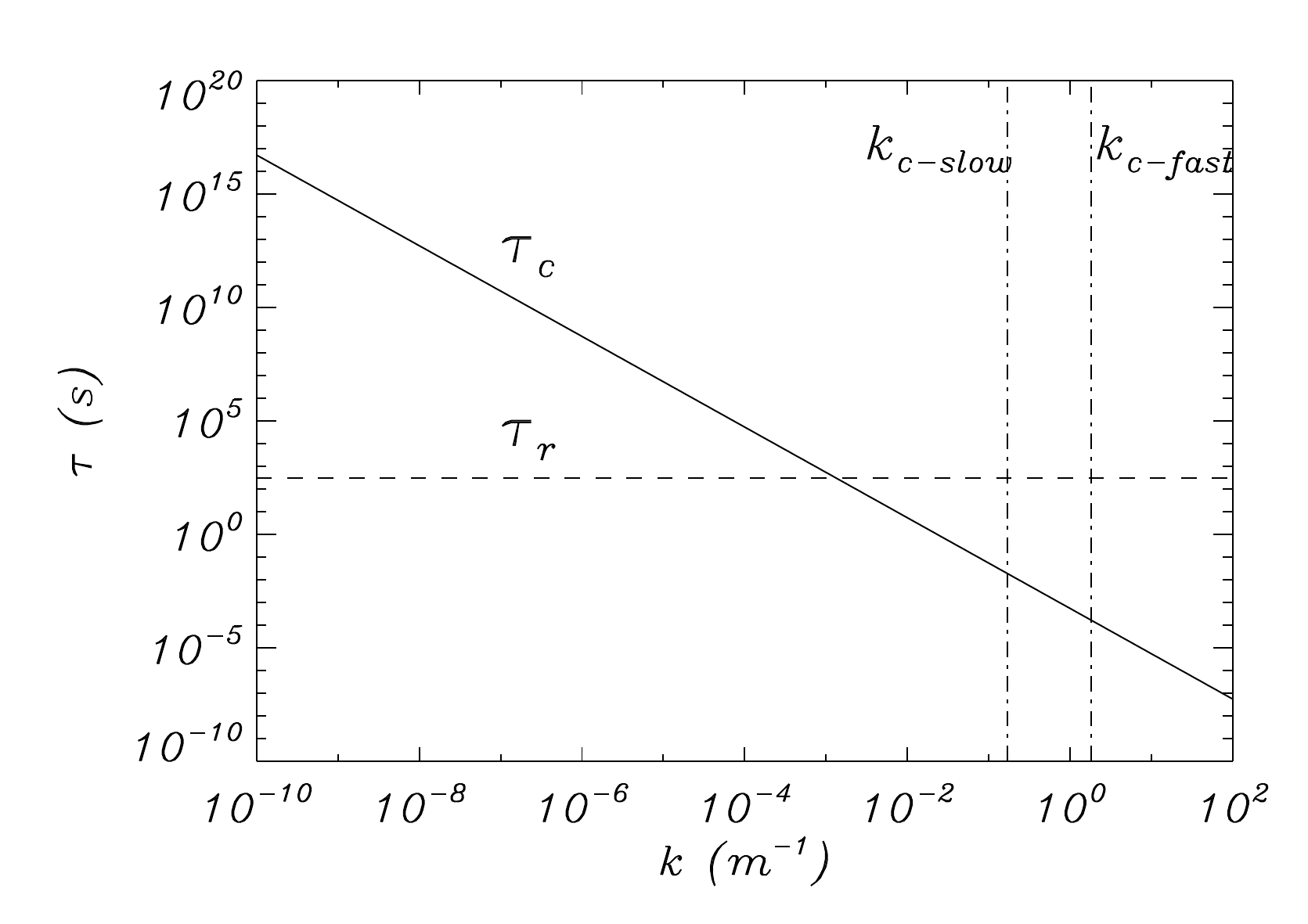}}
\caption{Time scales and critical wavelength in a uniform plasma with thermal conduction and radiative losses. Solid line: conduction time-scale, $\tau_c$; dashed line: radiative time-scale, $\tau_r$; dash-dotted lines: critical wavenumber that separates adiabatic from isothermal perturbations. The parameter values used correspond to a prominence plasma: $\rho=5\times10^{-11}$ kg m$^{-3}$, $T=8000$ K, $\tilde\mu=0.8$ (mean atomic weight), $B=10$ G. Moreover, parallel propagation ($\cos\theta=1$) has also been assumed, so that for other propagation angles the critical wavenumbers are displaced to the right with respect to the vertical dash-dotted lines shown here.}
\label{fig_nonadiab}
\end{figure}

Figure~\ref{fig_nonadiab} presents a plot of the characteristic conduction and radiative times together with the critical slow and fast mode wavenumbers marking the transition from the adiabatic to the isothermal regime. Now, let us consider a perturbation with tunable wavelength travelling in a magnetised, uniform medium in which thermal conduction and radiation can transport heat. Let us start with an extremely short wavelength, so that the wavenumber is larger than $k_c$ and the perturbation is isothermal (hence, we are at the right of Figure~\ref{fig_nonadiab}). Let us now increase the wavelength until the wavenumber becomes just smaller than $k_c$. Now perturbations are no longer isothermal, but given that thermal conduction is still more efficient than radiation, since it works in a very short time-scale (i.e. $\tau_c\ll\tau_r$), the perturbation is dominated by conduction. Next the wavelength is allowed to increase further and consequently the conduction time-scale increases proportionally to $L$. Under these conditions, conduction becomes less and less efficient and there is eventually a wavenumber (or length-scale) at which the conduction and radiation time-scales become equal; it corresponds to the position in Figure~\ref{fig_nonadiab} in which the solid and dashed lines cross. This length-scale, $L^*$, is obtained by imposing $\tau_c=\tau_r$ and with the help of Equations~(\ref{tau_c}) and (\ref{tau_r}) is

\begin{equation}
L^*=\frac{\gamma\kappa_\|}{\rho^2\chi T^{\alpha-1}}.
\end{equation}

\noindent Finally, for wavelengths larger than $L^*$ radiative losses dominate the energy budget of the plasma.

\begin{figure}[h]
\center{\mbox{
\includegraphics[height=3.8cm]{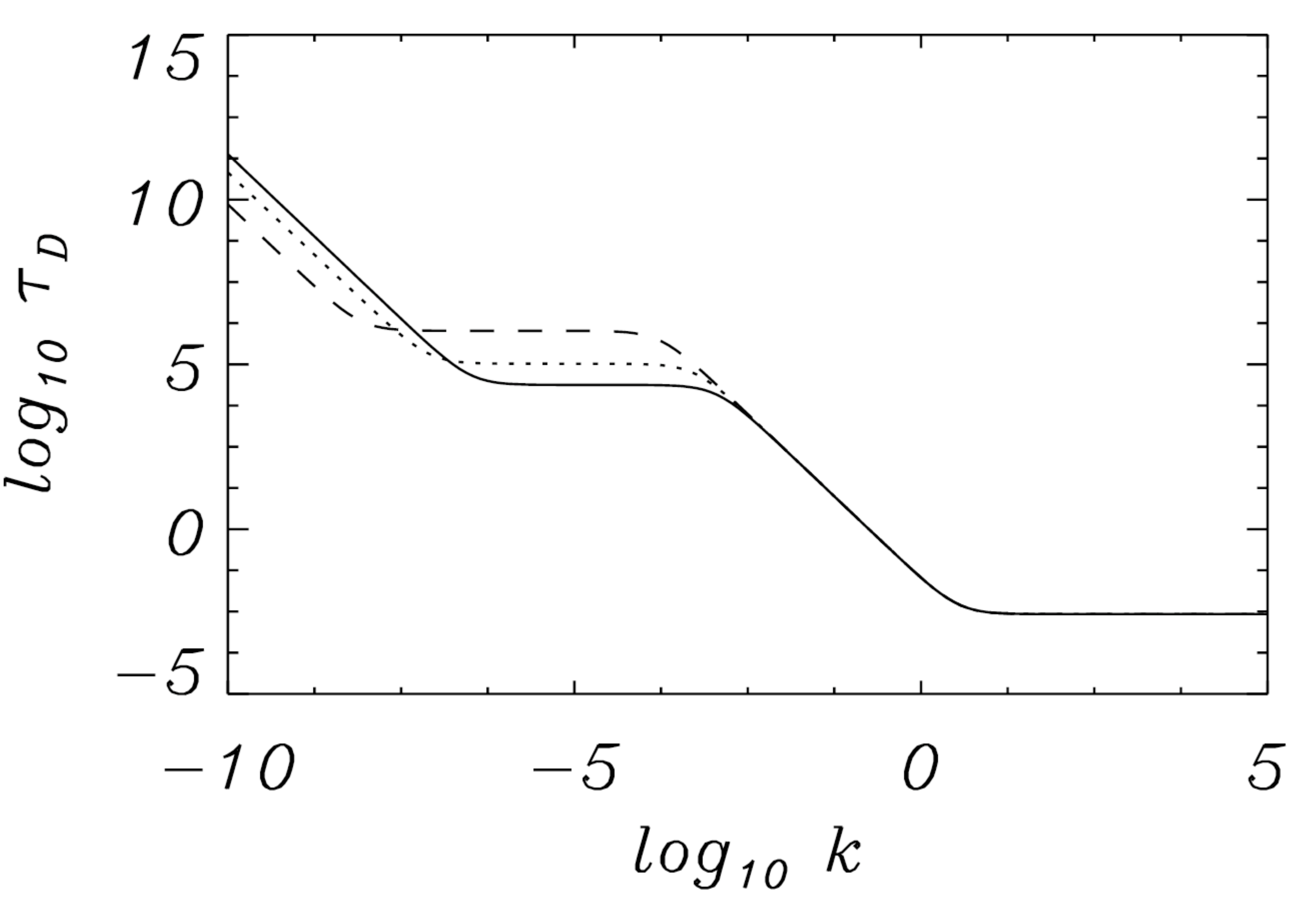}
\includegraphics[height=3.8cm]{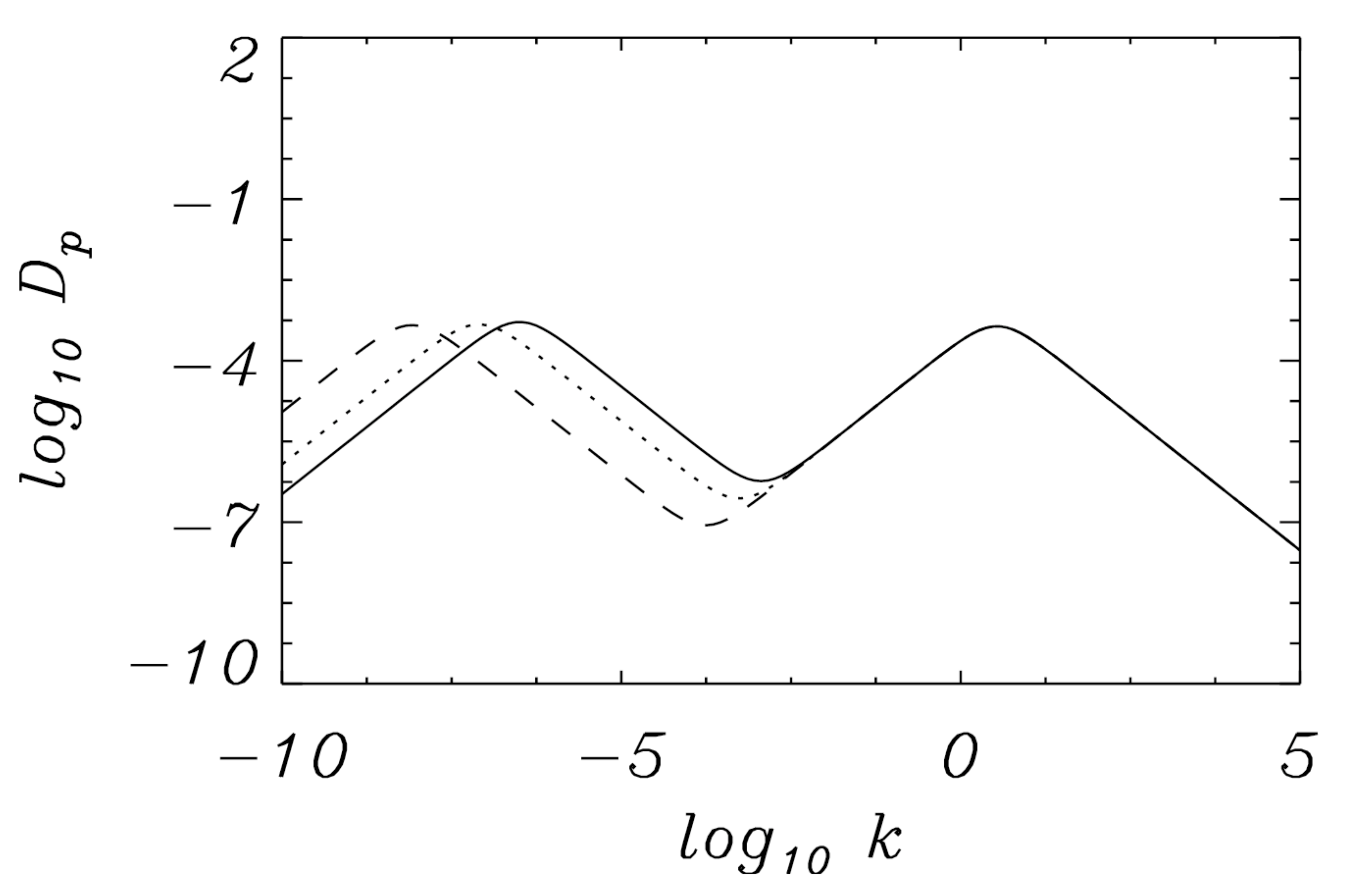}
}}
\caption{(a) Damping time and (b) damping per period, $D_p=2\pi\tau/P$, versus the wavenumber of the fast mode in an unbounded prominence medium with thermal conduction and radiative losses. The three curves correspond to different parametrisations of the prominence radiative losses. The wavenumber is given in m$^{-1}$. Figures taken from \cite{COB04}.}
\label{fig_nonadiab2}
\end{figure}

These formulas are useful to explain some features of the damping time (in the case of temporal damping of oscillations) or the damping length (in the case of spatial damping) obtained by \cite{IE93,COB04,TCOB05,CTOB06}. Consider, for example, the damping time and the damping per period ($D_p=2\pi\tau/P$) of the fast mode modified by radiative losses and thermal conduction in a uniform medium. Their variation with the wavenumber is displayed in Figure~\ref{fig_nonadiab2}, in which each of the three curves originates from a particular pair of values of $\chi$ and $\alpha$, the two parameters in the radiative loss function. In our discussion we concentrate on the solid line, although similar conclusions can be deduced for the other two (and also for the slow mode, which is not examined here). All curves in Figure~\ref{fig_nonadiab2} display three changes of slope whose origin can be partly explained on the basis of the transitions between different regimes studied above. For the paramater values used in this work, we have $k_{c-fast}=1.82$ m$^{-1}$ and $L^*=4770$ m, that corresponds to the wavenumber $k^*=2\pi/L^*=1.32\times 10^{-3}$ m$^{-1}$. The right-most change of slope in Figure~\ref{fig_nonadiab2} is around $\log k=0.5$ and coincides roughly with the value of $k_{c-fast}$, so it is caused by the transition between the isothermal and the conduction-dominated regimes. In addition, the intermediate change of slope is around $\log k=-3$ and its agreement with $k^*$ confirms that it corresponds to the transition from the conduction-dominated to the radiation-dominated regimes. Therefore, the shift from one regime to another leaves its imprint in the variation of $\tau$ and $\tau/P$ with respect to the wavenumber. We also note that, since the difference between the three curves of Figure~\ref{fig_nonadiab2} lies in the radiative loss function used, they overlap in the conduction dominated regime.

The main conclusions of \cite{COB04,TCOB05,CTOB06} regarding the damping of prominence oscillations are that the slow mode is strongly attenuated for wavenumbers within the range of observed values (i.e. for $k$ in the range $10^{-8}-10^{-6}$ m$^{-1}$), while the fast mode is not so strongly damped and its attenuation rate is too small compared with the values from observations. Moreover, the considered approximations for optically thin or thick plasmas can give very different attenuation properties and all the considered heating mechanisms are of no relevance.

\subsection{Radiative losses and thermal conduction in a structured medium: prominence slab}\label{nonadiab_slab}

The influence of the surrounding corona on the damping of magnetoacoustic oscillations via thermal conduction and radiative losses has been studied by \cite{SOB07a,SOB09}.

\begin{figure}[h]
\center{\mbox{
\includegraphics[height=3.8cm]{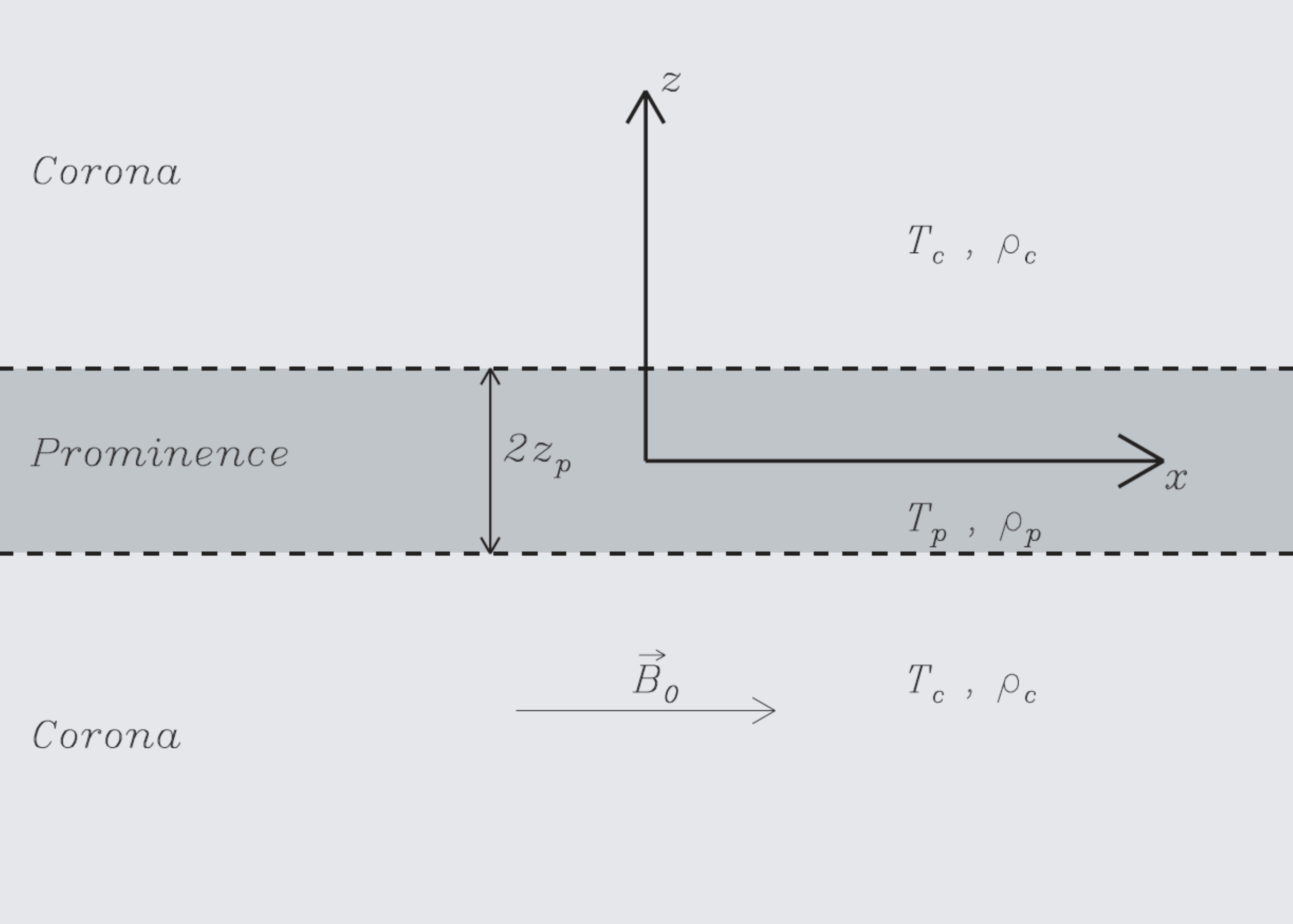}
\includegraphics[height=3.8cm]{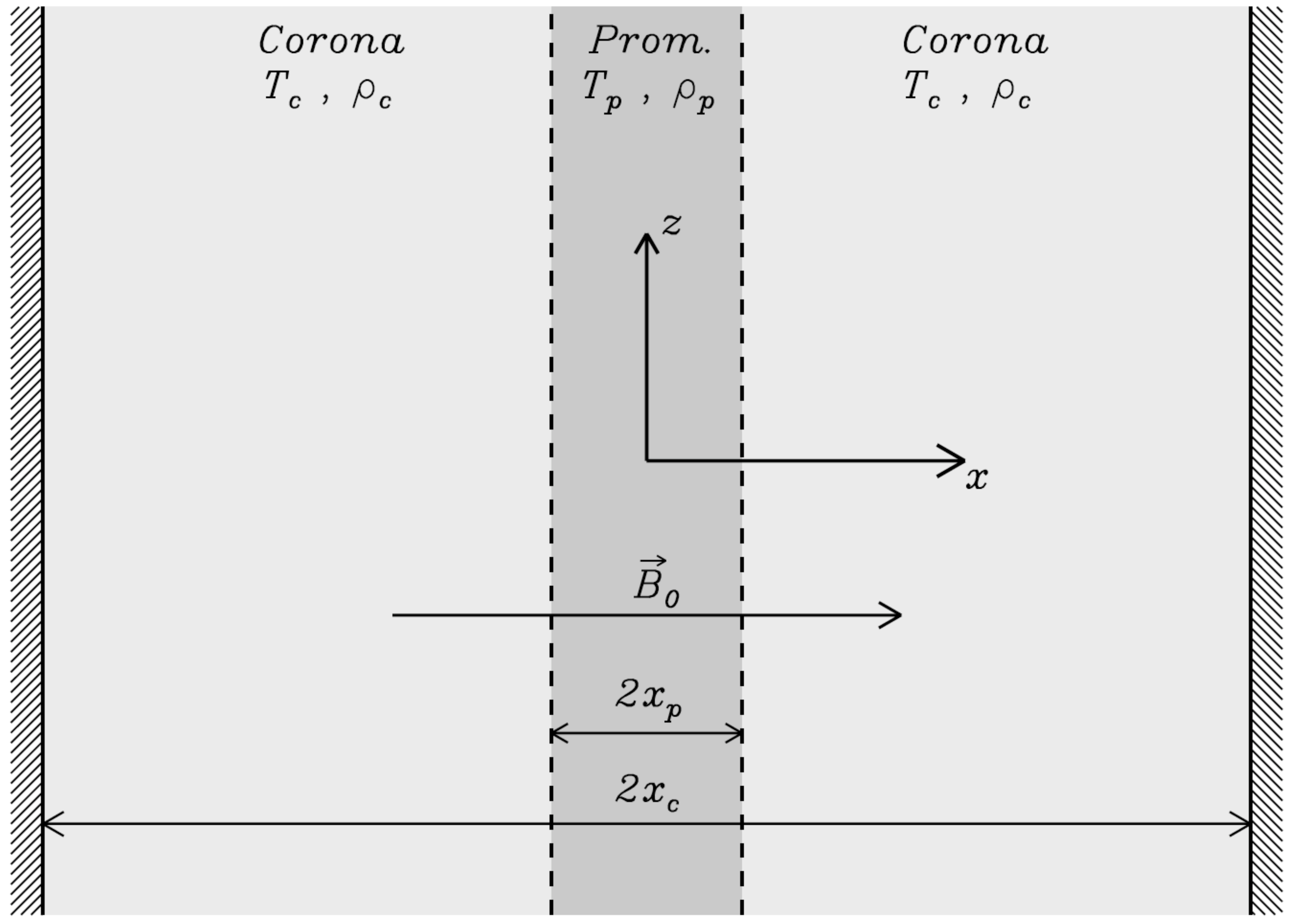}
}}
\caption{Sketch of a prominence slab (temperature $T_p$ and density $\rho_p$) embedded in the solar corona (temperature $T_c$ and density $\rho_c$). Left: longitudinal magnetic field, parallel to the prominence length. Right: transverse magnetic field, perpendicular to the prominence length. Figures taken from \cite{SOB07a,SOB09}.}
\label{fig_struct_along}
\end{figure}

\begin{figure}[h]\sidecaption
\resizebox{0.5\hsize}{!}{\includegraphics*{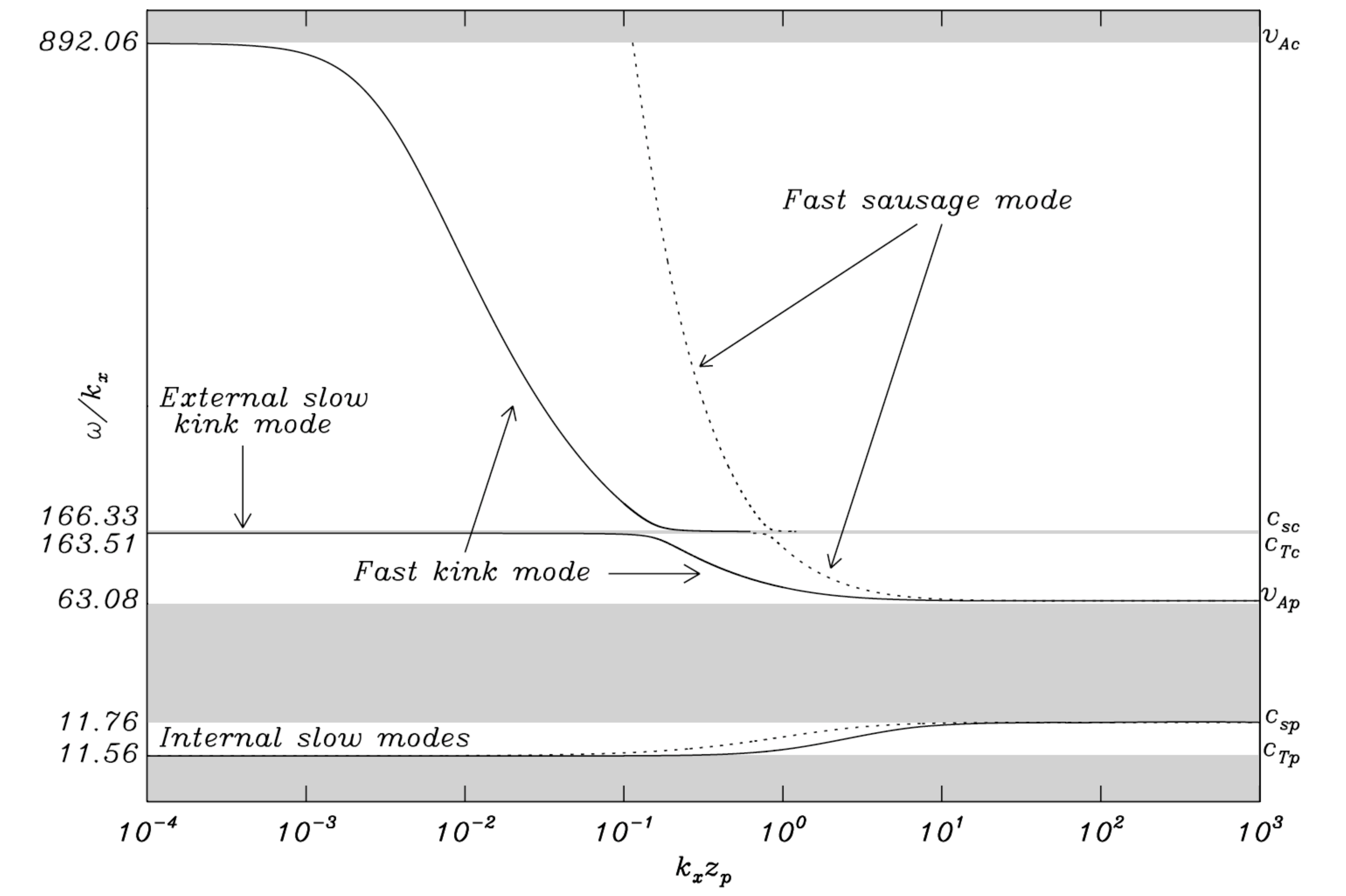}}
\caption{Phase speed (in km s$^{-1}$) versus the dimensionless longitudinal wavenumber of the normal modes of a prominence-corona system with longitudinal magnetic field. Solid lines denote kink modes and dotted lines denote sausage modes. Only the fundamental modes are plotted. The shaded zones are forbidden regions where evanescent-like solutions in the corona do not exist. Note that the vertical axis is not drawn to scale. Figure taken from \cite{SOB07b}.}
\label{fig_modes_along}
\end{figure}

In \cite{SOB07a} the filament is taken as a plasma slab embedded in the solar corona and with a uniform magnetic field parallel to the filament axis; see Figure~\ref{fig_struct_along}a. Both the prominence plasma and the coronal plasma are uniform and threaded by the same magnetic field. To understand the results of \cite{SOB07a} it is necessary to describe the main properties of the linear, adiabatic normal modes of such a configuration, that have been studied by \cite{ER82,JR92a,SOB07b}. In the latter work the solutions are classified in three types (Figure~\ref{fig_modes_along}): fast, internal slow and external slow modes and it is concluded that the perturbations associated to internal slow modes are essentially confined to the prominence slab, so that these waves are hardly influenced by the coronal environment. On the other hand, fast modes have tails that penetrate in the corona and as a result coronal conditions are important in determining the features of these waves. Moreover, the confinement of fast modes becomes poorer for small values of the wavenumber in the direction of the filament axis. Regarding external slow modes, they produce almost negligible perturbed amplitudes inside the cold slab and thus seem uninteresting in the context of prominence oscillations.

\begin{figure}[h]\sidecaption
\resizebox{0.5\hsize}{!}{\includegraphics*{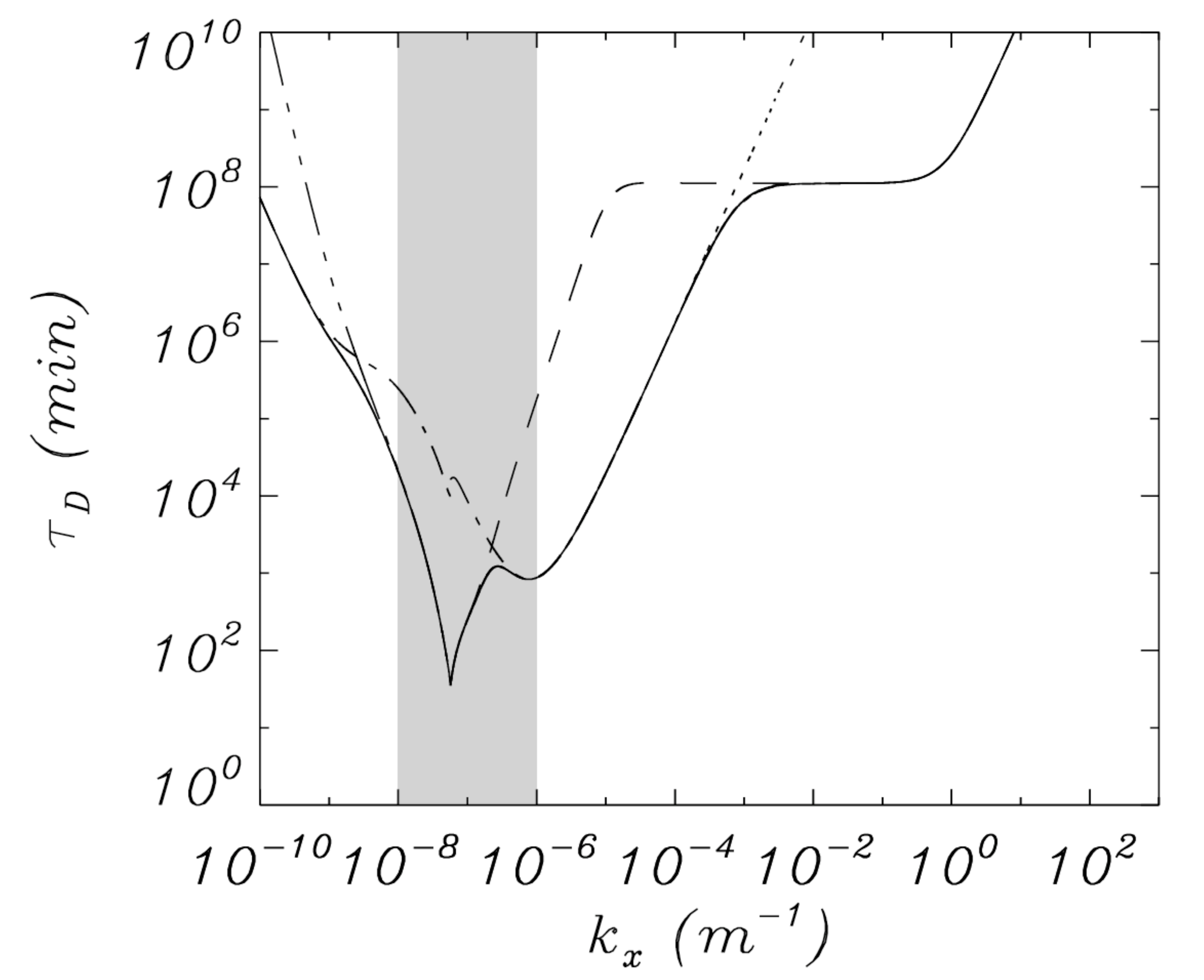}}
\caption{Damping time versus the longitudinal wavenumber for the fundamental fast kink mode in a prominence slab with longitudinal magnetic field. Different linestyles represent the results obtained when different non-adiabatic mechanisms are omitted: all mechanisms considered (solid line), prominence conduction eliminated (dotted line), prominence radiation eliminated (dashed line), coronal conduction eliminated (dot-dashed line) and coronal radiation eliminated (three dot-dashed line). The shaded region indicates the interval of observed wavelengths. Figure taken from \cite{SOB07a}.}
\label{fig_fast_mode_along}
\end{figure}

\cite{SOB07a} included the thermal conduction, radiative losses and plasma heating terms in the energy equation and studied the damping properties of the fast and internal and external slow modes. They obtained values of $\tau/P$ in agreement with observations both for the internal slow and the fast modes; this last result implies that the presence of the surrounding corona modifies the damping properties of fast waves. Such as expected, the damping rate of the internal slow mode is not modified by the inclusion of the corona in the model and it only depends on the value of the wavenumber in the direction parallel to the magnetic field. The attenuation of this mode can be recovered by inserting the parallel wavenumber, $k_x$, in the expression $\omega=k_xc_s$ and by substituting $c_s$ by a ``modified sound speed'', $\Lambda$, whose square is given by

\begin{equation}\label{nonadiab_sound_speed}
\Lambda^2=\frac{c_s^2}{\gamma}\,\frac{(\gamma-1)\left(\frac{T}{p}\kappa_\|k_x^2+\omega_T-\omega_\rho\right) + i\gamma\omega}{(\gamma-1)\left(\frac{T}{p}\kappa_\|k_x^2+\omega_T\right) + i\omega},
\end{equation}

\noindent where $\omega_\rho$ and $\omega_T$ are determined from the partial derivatives of the heat-loss function with respect to density and temperature, respectively. The damping time of the fast kink mode, however, is severely influenced by the addition of the coronal medium. This can be clearly appreciated by comparing the damping time in a uniform medium (Figure~\ref{fig_nonadiab2}a) and in a structured medium with longitudinal magnetic field (Figure~\ref{fig_fast_mode_along}). The reason for such a strong modification of the fast kink mode damping time is, firstly, that this mode couples with the external slow mode, whose properties are essentially dictated by the coronal plasma. And secondly, apart from this coupling the coronal medium has a direct influence because of the presence of coronal tails in the perturbed variables. In \cite{SOB07a} a close examination of the role of the corona in the damping of the fast mode was done by removing one of the important non-adiabatic mechanisms (thermal conduction or radiation) either in the filament or in the corona and then computing the period and damping time. The relevance of each mechanism is revealed by a discrepancy between the damping time computed with and without this mechanism (Figure~\ref{fig_fast_mode_along}). The conclusion of this study is that coronal mechanisms govern the attenuation of the fast modes for wavenumbers $k_x\lesssim 10^{-7}$~m$^{-1}$, whereas prominence mechanisms prevail for $k_x\gtrsim 10^{-7}$ m$^{-1}$. In each of these ranges radiation is dominant in the low-$k_x$ range and thermal conduction is dominant in the high-$k_x$ range, such as found in Section~\ref{non-adiab-unif-medium} for a uniform medium. More specifically, coronal radiation controls the attenuation for $k_x\lesssim 10^{-9}$ m$^{-1}$, coronal conduction does so for $10^{-9}$ m$^{-1}$ $\lesssim k_x\lesssim 10^{-7}$ m$^{-1}$, while prominence radiation and conduction are the most important non-adiabatic mechanisms for $10^{-7}$ m$^{-1}$ $\lesssim k_x\lesssim 10^{-3}$ m$^{-1}$ and $k_x\gtrsim 10^{-3}$ m$^{-1}$, respectively. Hence, in the range of observed wavelengths both thermal conduction in the corona and radiation from the prominence gas are important to explain the reported damping times. In Section~\ref{non-adiab-unif-medium} the characteristic time-scales of thermal conduction and radiation in a uniform medium were invoked to explain the transition from the radiation-dominated wavenumber range to the conduction-dominated wavenumber range. The same idea can be applied to explain this transition in the present, structured configuration both for the prominence and for the corona. In the case of the prominence, the transition value is the same as before, namely $k^*=1.32\times 10^{-3}$ m$^{-1}$, whereas for the corona it is $k^*=3.42\times 10^{-9}$~m$^{-1}$. These numbers coincide well with the transitions at which different mechanisms dominate in Figure~\ref{fig_fast_mode_along}. Moreover, the wavenumber at which the transition between the ranges of dominance of coronal conduction and prominence radiation takes place can be derived from the equality of $\tau_c$ in the corona and $\tau_r$ in the prominence. This calculation yields $k=1.32\times 10^{-7}$ m$^{-1}$, also in good agreement with the value from Figure~\ref{fig_fast_mode_along}. More details can be found in \cite{SOB07a}.

In \cite{SOB09} a prominence-corona configuration with an equilibrium magnetic field perpendicular to the prominence slab is considered; see Figure~\ref{fig_struct_along}b. Such a structure can support fast and slow modes of internal or external character, depending on whether their properties are governed mainly by the prominence or the corona, respectively. Moreover, there is another type of fast and slow mode, called string by \cite{JR92b} and hybrid by \cite{OBHP93}, that is simultaneously internal and external. \cite{SOB09} find that slow modes are efficiently damped by the considered non-adiabatic mechanisms (thermal conduction and radiative losses), such as found by \cite{SOB07a} for a perpendicular orientation of the magnetic field. It is also shown that prominence radiation dominates the attenuation of internal slow modes and that prominence radiation and coronal conduction together dominate that of the hybrid slow mode. This result for the hybrid mode is coherent with the fact that its perturbations achieve large amplitudes both in the prominence and the corona, so one expects that the most relevant damping mechanisms of each medium govern together the attenuation of the hybrid mode. On the other hand, fast modes are very poorly damped in the present configuration, which implies that the assumed magnetic field orientation is decisive for these modes. A remarkable feature of fast modes is that they can become unstable under certain circumstances. Thermal conduction allows energy transfer between the prominence slab and the coronal medium. Prominence radiation has an essential role in dissipating the extra heat injected from the corona and stabilises the oscillations. Nevertheless, thermal instabilities appear if radiative losses from the prominence plasma are omitted or significantly reduced (e.g. caused by an increase of the optical thickness) since the plasma cannot dissipate the extra injected heat in an efficient way. Given that different parametrisations of the radiative loss function have been given in the literature \citep{H74,RTV78,MPR79} and that they correspond to different optical thicknesses, the relevance of the optical thickness in the stability of fast waves is not firmly established yet.

\subsection{Attenuation by radiative losses and thermal conduction in the presence of flow}

\begin{figure}[h]
\vspace{-2.8cm}\center{\mbox{
\includegraphics[height=10cm]{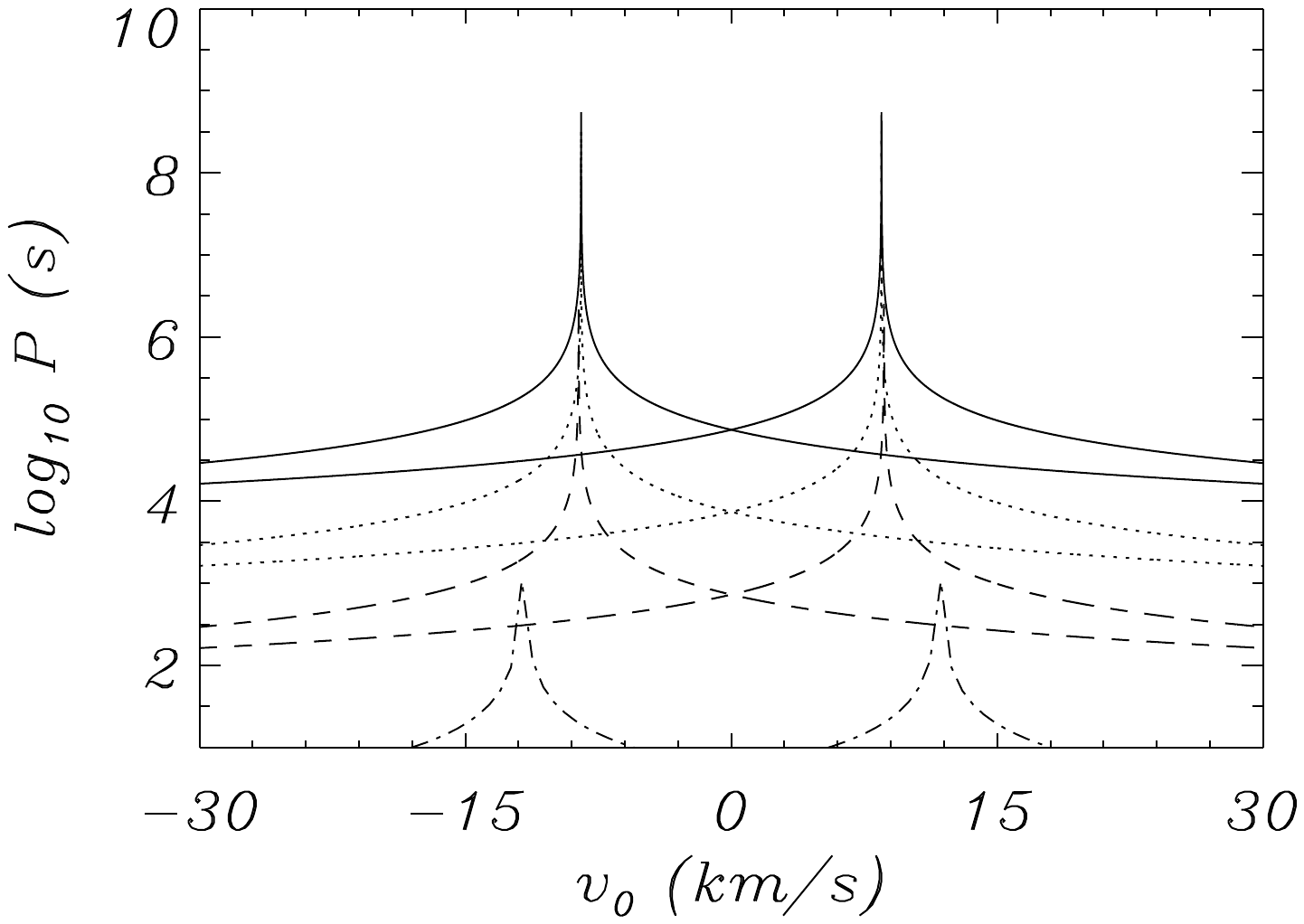}
\hspace{-2cm}\includegraphics[height=10cm]{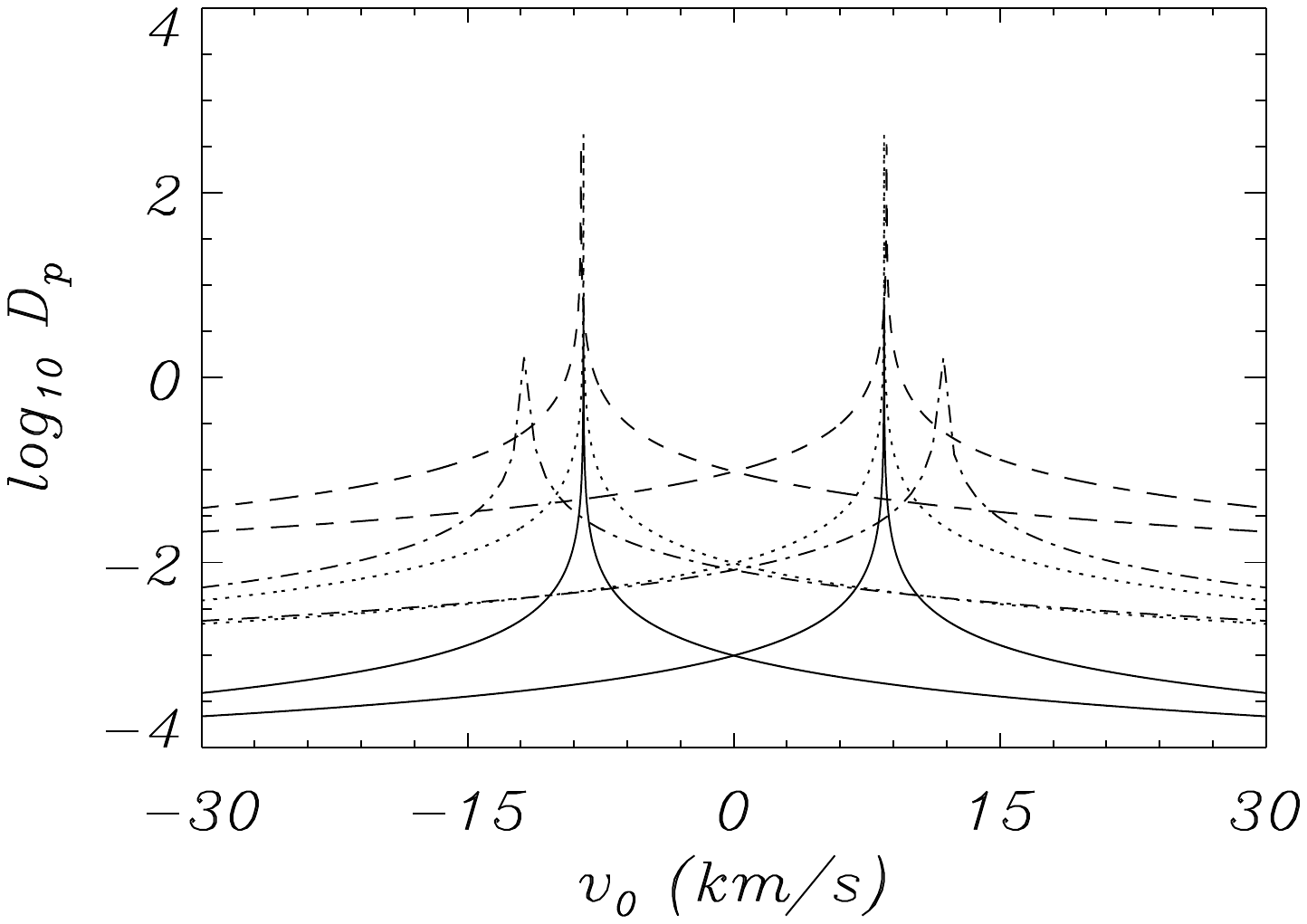}
}}
\vspace{-4cm}\caption{(a) Period and (b) damping per period, $D_p=2\pi\tau/P$, versus the flow speed for slow waves in a uniform plasma with flowing motions and subject to radiative losses and thermal conduction. Solid, dotted, dashed and dash-dotted lines correspond to wavenumbers $k_x=10^{-8},10^{-7},10^{-6},10^{-4}$ m$^{-1}$, respectively. The inclusion of flow in the model gives rise to the presence of two slow modes for each wavenumber when the flow speed is non-zero. Figures taken from \cite{COB09}.}
\label{fig_nonadiab_flow}
\end{figure}

Material flows are common features in solar prominences \citep{ZEM98,LEW03,LEvdVvN07}. \cite{COB09} investigated the attenuation of the slow and thermal waves by radiative losses and thermal conduction in a uniform medium with flow. In this work the fast mode is ignored because these non-adiabatic effects do not contribute in a significant way to its attenuation \citep{COB04}; slow and thermal waves are investigated in the limit of field-aligned propagation. \cite{COB09} find that the period and damping per period can show a strong dependence on $v_0$, the flow speed, for certain values of this parameter; Figure~\ref{fig_nonadiab_flow} presents their results for the slow mode. They also find that the greatest period and damping per period of slow waves is obtained for flow speeds close to the real part of the non-adiabatic sound speed, which in the presence of flow is deduced from Equation~(\ref{nonadiab_sound_speed}) with $\omega$ substituted by $\omega-k_xv_0$. Regarding the thermal mode, which in the absence of flow does not propagate, it is found that it becomes propagating for $v_0\neq 0$. Therefore, when flows and time damped oscillations are observed in a prominence/filament, this wave should be also taken into account as potentially responsible of the observed damped oscillations. These authors remark that it may not be possible to determine whether an observed period and damping time are associated to a slow wave or a thermal wave in a flowing material and that a possibility to determine the origin of such wave is to resort to the temperature perturbation, which should be rather large for the thermal wave and much smaller for the slow wave.

\subsection{Radiative losses and thermal conduction in a structured medium: cylindrical prominence thread}

\begin{figure}[h]
\center{\mbox{
\includegraphics[height=3.5cm]{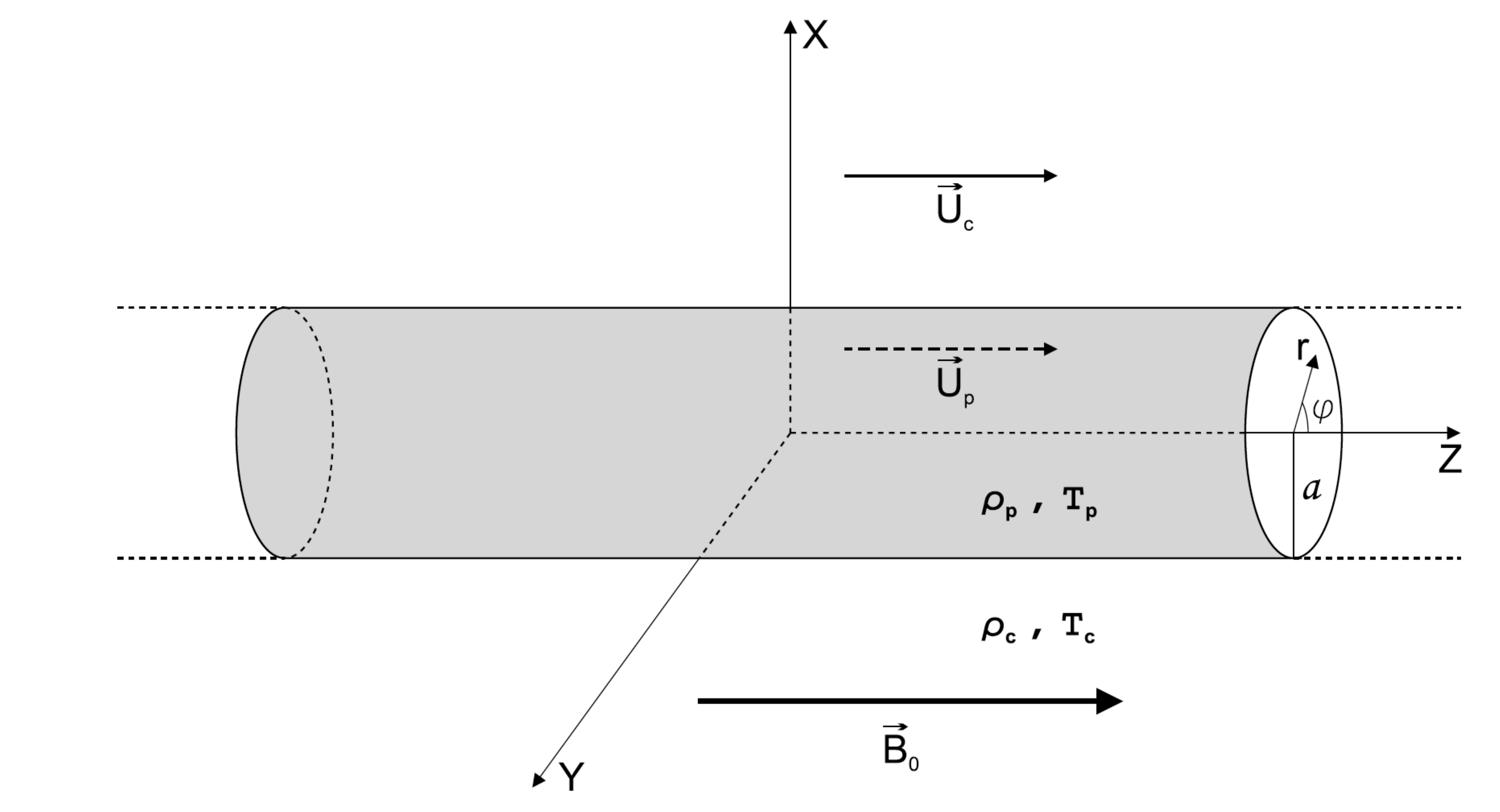}
\includegraphics[height=3.3cm]{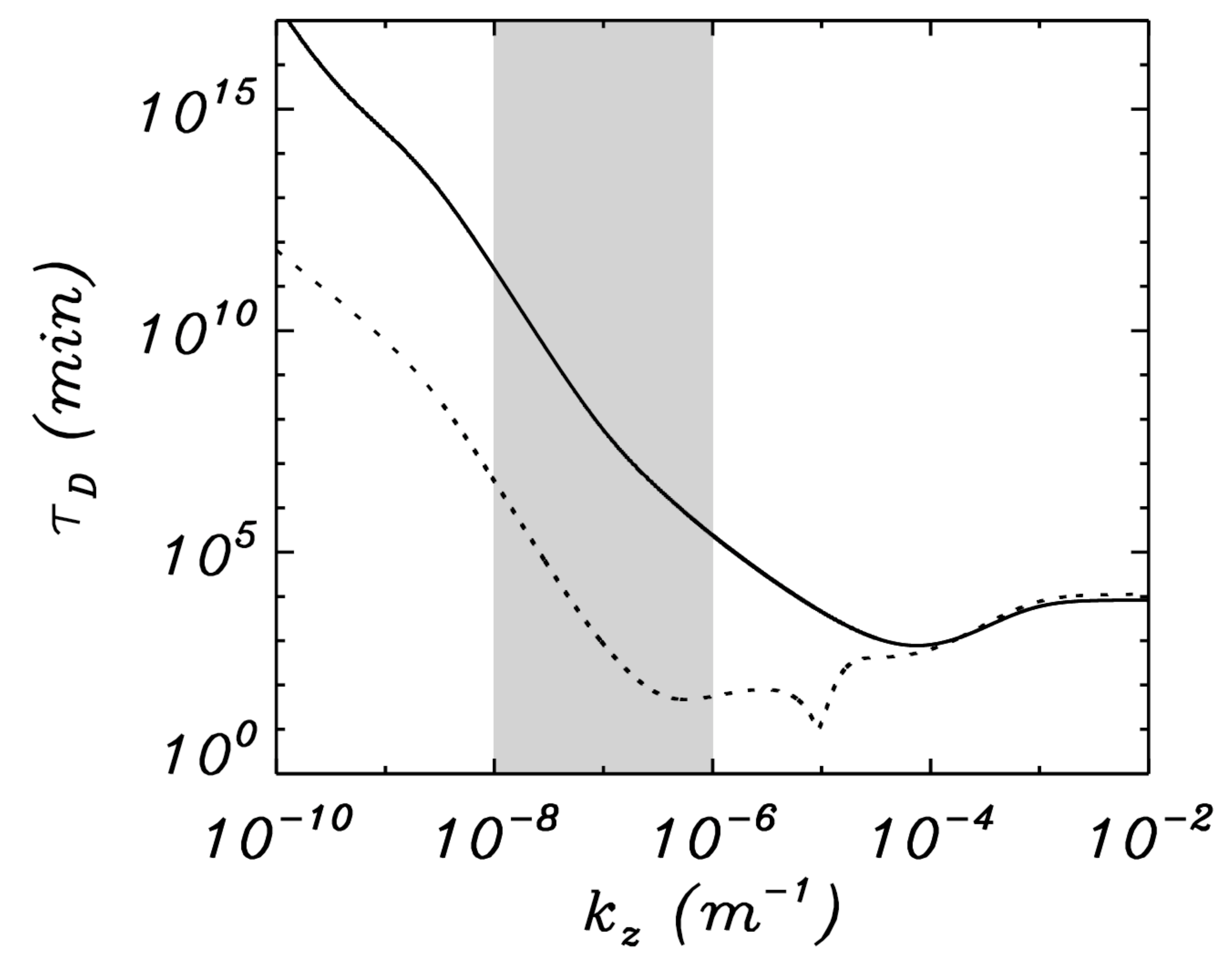}
}}
\caption{(a) Sketch of the thread model used by \cite{SOB08} to study the damping of oscillations caused by thermal conduction and radiative losses in the presence of mass flow. (b) Damping time of the fast kink mode in a cylindrical thread (solid) and in a thin slab with longitudinal magnetic field (dotted). In both cases the flow velocity is zero. The shaded area indicates the interval of observed wavelengths. Figures taken from \cite{SOB08}.}
\label{fig_nonadiab_thread}
\end{figure}

Instead of considering the attenuation of perturbations in a full-prominence configuration, as in Section~\ref{nonadiab_slab}, \cite{SOB08} restrict themselves to a single thread. Their model consists of a homogeneous, infinite cylinder representing the thread embedded in an unbounded corona; see Figure~\ref{fig_nonadiab_thread}a. Two additional ingredients of the model are a uniform magnetic field both inside and outside the thread and flow motions parallel to the magnetic field in both regions. Thermal conduction and radiative losses are taken into account as damping mechanisms, and the combined effect of these non-ideal physics and the steady flow on the attenuation of oscillations is assessed. In this work it is found that, in the absence of flow, slow modes are efficiently damped by non-adiabatic effects, while fast kink waves are in practice not attenuated, since their damping times are much larger than typical lifetimes of filament threads. Therefore, the damping by non-adiabatic mechanisms of kink oscillations is much less efficient in this cylindrical case than in the slab geometry (Figure~\ref{fig_nonadiab_thread}b). In the presence of flow it is well-known that the symmetry between waves propagating parallel or antiparallel to the flow is broken. The thermal mode behaves as a wave that propagates parallel to the flow, and its motions are mainly polarized along the longitudinal direction. Nevertheless, this oscillatory behaviour is not likely to be observed in practice due to its quick damping. The damping time of both slow and thermal waves is not
affected by the flow. On the contrary, for realistic values of the flow velocity, the larger the flow, the larger the attenuation of parallel fast kink waves, whereas the contrary occurs for antiparallel fast kink solutions. Nevertheless, this effect is not enough to obtain realistic damping times in the case of fast kink modes.

\subsection{Partial ionisation effects in a uniform medium}\label{ionneutral}

Chromospheric and prominence plasmas are not fully ionised and so several new effects are present in comparison to a fully ionised plasma. For example, electric charges are frozen to magnetic field lines, but neutrals are not and thus the neutral and ionised fractions of the plasma behave differently. Collisions between neutrals, on one hand, and electrons and ions, on the other hand, arise and consequently a modified Ohm's law is obtained. As a consequence, the main outcome of the interaction between neutrals and charged particles is the presence of enhanced Joule heating and enhanced magnetic diffusion. These ideas were presented by \cite{P56} in connection with the heating of the solar chromosphere and corona and have also been invoked to explain the dynamics of chromospheric spicules \citep[e.g.][]{H92,JE02} and of flux tubes emerging from the solar interior \citep{AHL07}.

\begin{figure}[h]\sidecaption
\resizebox{0.5\hsize}{!}{\includegraphics*{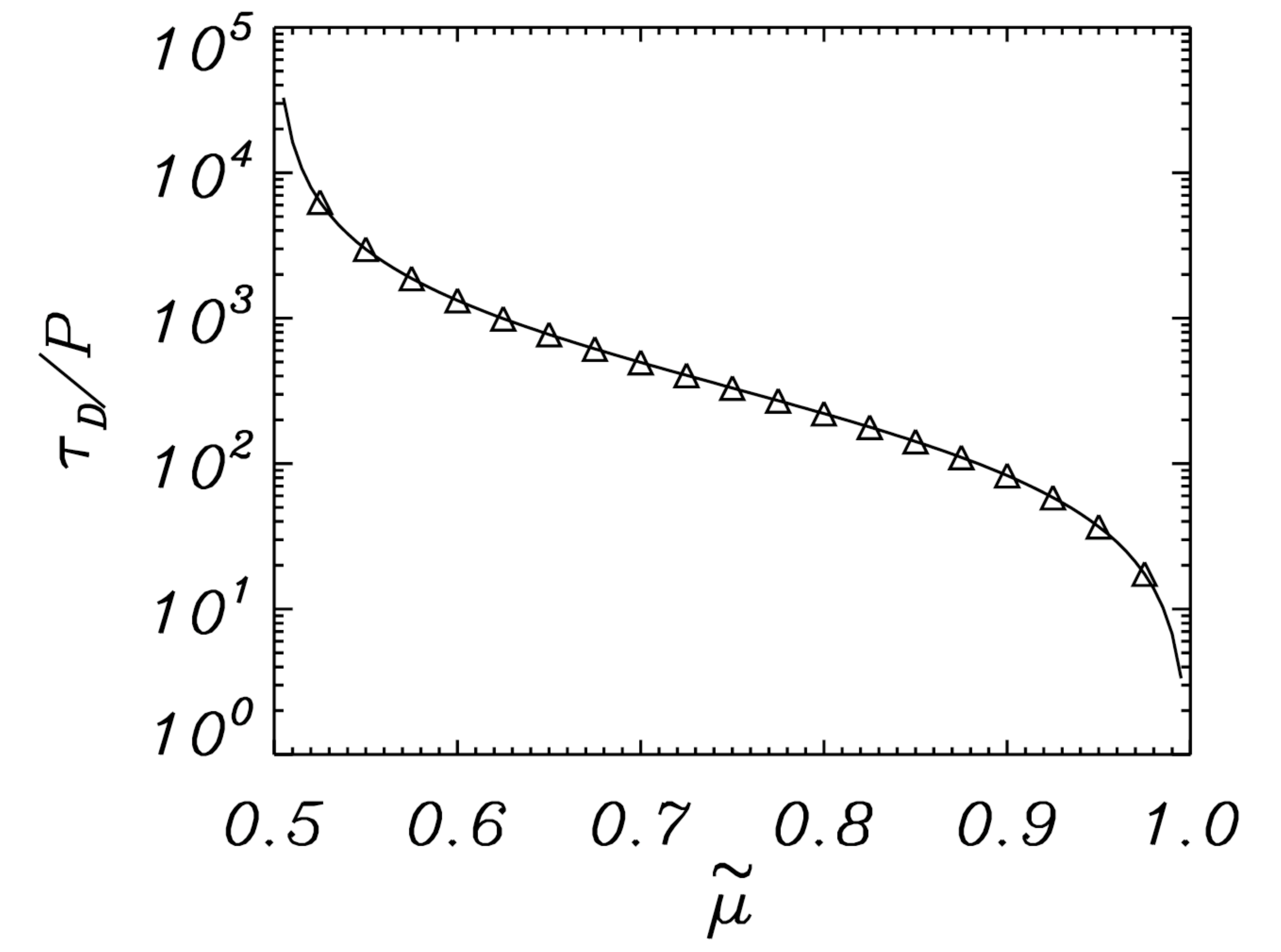}}
\caption{Damping rate versus the ionisation fraction, $\tilde\mu=1/(2-\xi_n)$, for the fast mode in a uniform, partially ionised plasma. The solid line represents the numerical solutions of the full dispersion relation while the triangles represent the results obtained by solving the aproximate Equation~(\ref{eq_ppi}). Figure taken from \cite{FOBK07}.}
\label{fig_pip}
\end{figure}

The influence of partial ionisation of the prominence plasma on the properties of the fast and slow modes has been investigated by \cite{FOBK07}. In this work the one-fluid MHD equations for a partially ionised plasma made of electrons, protons and neutral hydrogen are derived. Next, a uniform medium with a straight magnetic field is considered. The linear regime is assumed and this results in the vanishing of the Joule heating term in the energy equation, so that the corresponding physical effect is absent. One may conclude that in more realistic configurations dissipation may be stronger than that found by \cite{FOBK07}. The main result of this work is that the fast mode can be strongly damped (in agreement with observations) for small values of the ionisation fraction (Figure~\ref{fig_pip}), whereas the slow mode is hardly attenuated. The mechanism responsible for the damping of the fast mode is Cowling's diffusion, that vanishes in a fully ionised plasma. An analytical approximation for the fast mode damping rate,

\begin{equation}\label{eq_ppi}
\frac{\tau}{P}=\frac{1-\xi_n}{\xi_n^2}\rho^{1/2}B^{-1}k^{-1},
\end{equation}

\noindent is derived by \cite{FOBK07}, with $\xi_n$ the relative density of neutrals, $\rho$ the gas density, $B$ the magnetic field strength and $k$ the modulus of the wavenumber. Such a simple expression can be of practical use in the interpretation of observations. It indicates that the damping is stronger, i.e. $\tau/P$ is smaller, for almost neutral plasmas with low density, strong magnetic field and for short wavelengths.

\subsection{Radiative losses and thermal conduction in a uniform, partially ionised medium}

\begin{figure}[h]
\center{\mbox{
\includegraphics[height=3.8cm]{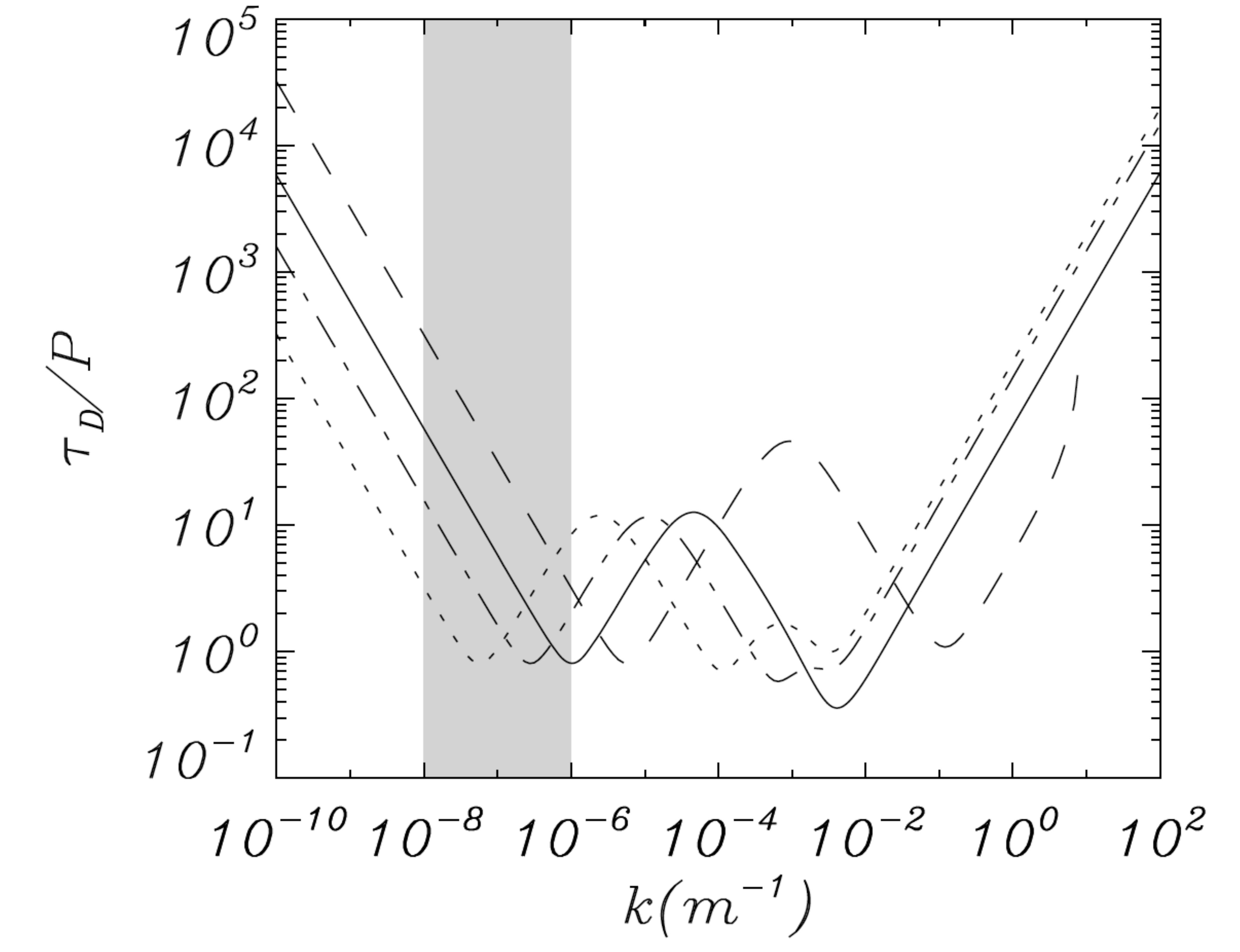}
\includegraphics[height=4cm]{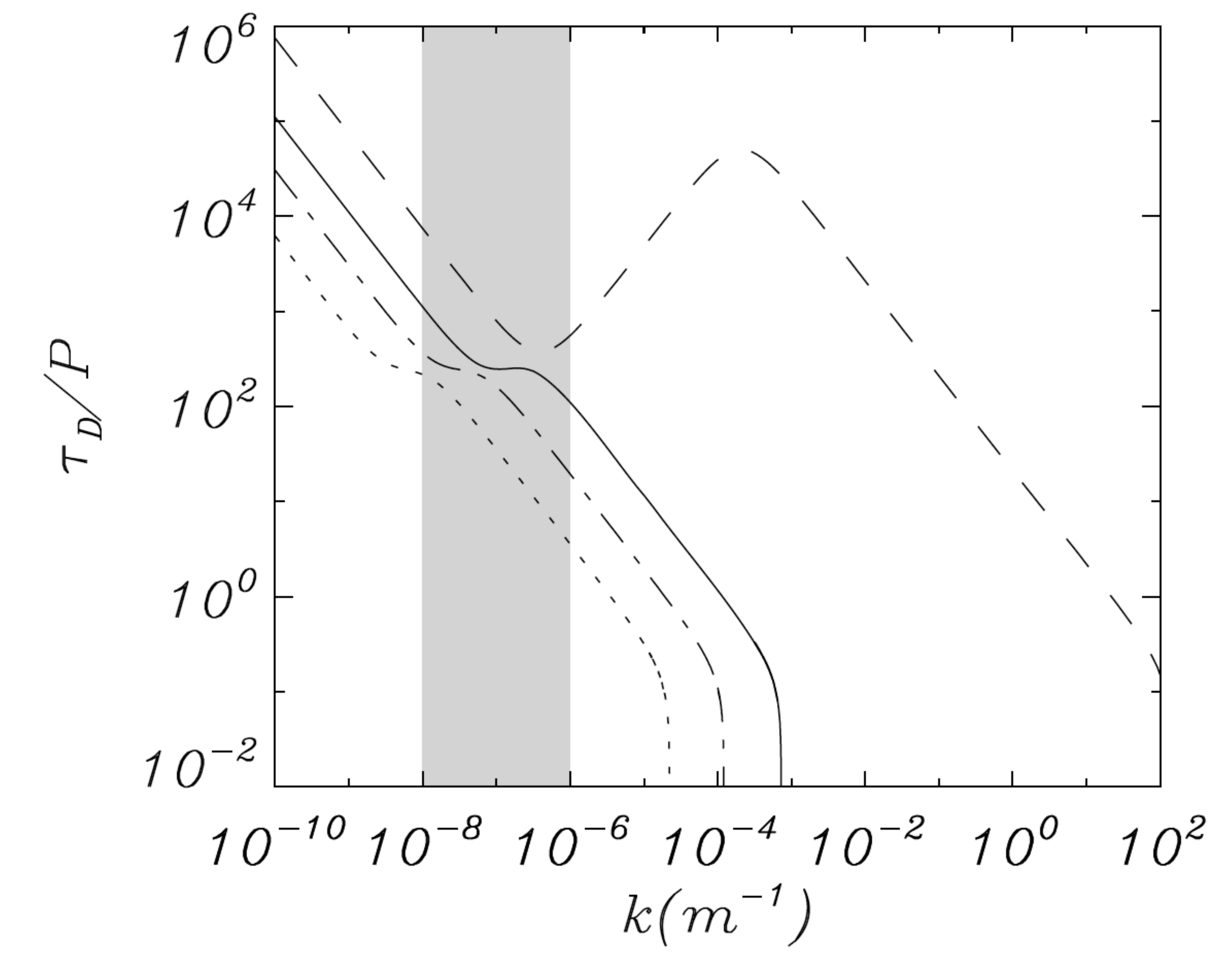}
}}
\caption{Ratio of the damping time to the period as a function of the wavenumber, $k$, for the (a) slow and (b) fast waves in a partially ionised plasma with thermal conduction and radiative losses. Different line styles correspond to different ionisation fractions: $\tilde\mu=0.5$ (dashed), $\tilde\mu=0.8$ (solid), $\tilde\mu=0.95$ (dash-dotted) and $\tilde\mu=0.99$ (dotted). The shaded region corresponds to the interval of observed wavelengths in prominences. Figures taken from \cite{FOB08}.}
\label{fig_pip_nonadiab}
\end{figure}

A natural extension of \cite{COB04}, concerned with the attenuation of fast and slow waves by thermal conduction and radiation, and \cite{FOBK07}, concerned with the attenuation of the same waves by ion-neutral effects, is to consider together a partially ionised plasma with thermal conduction and radiative losses. This task was performed by \cite{FOB08}, who derived the equations governing the dynamics of such medium. A particularly interesting issue is the different conduction properties of electrons (mainly along magnetic field lines) and neutrals (isotropic), which implies that the contribution of the neutral species to the thermal conduction must be taken into account separately following \cite{P53,IM90}. \cite{FOB08} find that the period of the magnetoacoustic waves remains basically the same as in the ideal case and that their model gives values of the ratio of the damping time to the period similar to the ones obtained in observations for the three wave modes (fast, slow and Alfv\'en). In addition, the inclusion of non-adiabatic terms in the partially ionised set of equations increases the damping of fast and slow waves in the interval of observed wavelegths as compared with the results obtained for a non-adiabatic fully ionised plasma \citep{COB04} and an adiabatic partially ionised plasma \citep{FOBK07}. For slow waves, the minima of the ratio of $\tau/P$, corresponding to attenuation maxima, are displaced towards longer wavelengths as compared to when only thermal conduction and radiative losses are included in the model. An increase of the neutral portion in the plasma produces a displacement of these ranges of maximum damping to longer wavelengths (Figure~\ref{fig_pip_nonadiab}a). Radiative losses are dominant for long wavelengths \citep[just as found by][]{COB04}, while the rest of the wavenumber interval is dominated by thermal conduction and ion-neutral collisions. Regarding fast waves, radiation is the dominant damping mechanism for long wavelengths, while in the rest of the considered wavenumber interval the attenuation is dominated by the effect of ion-neutral collisions. An important result is that fast waves only exist for wavenumbers smaller than a critical value that depends on the ionisation fraction (see Figure~\ref{fig_pip_nonadiab}b). In spite of this, the critical wavenumber is large in comparison with the typical wavenumbers of waves in prominences. Finally, for typical prominence temperature values, the contribution of electrons to thermal conduction is negligible in front of the contribution of neutrals.

\section{Damping by resonant absorption in a prominence thread}\label{resabs}

\begin{figure}[h]\sidecaption
\resizebox{0.5\hsize}{!}{\includegraphics*{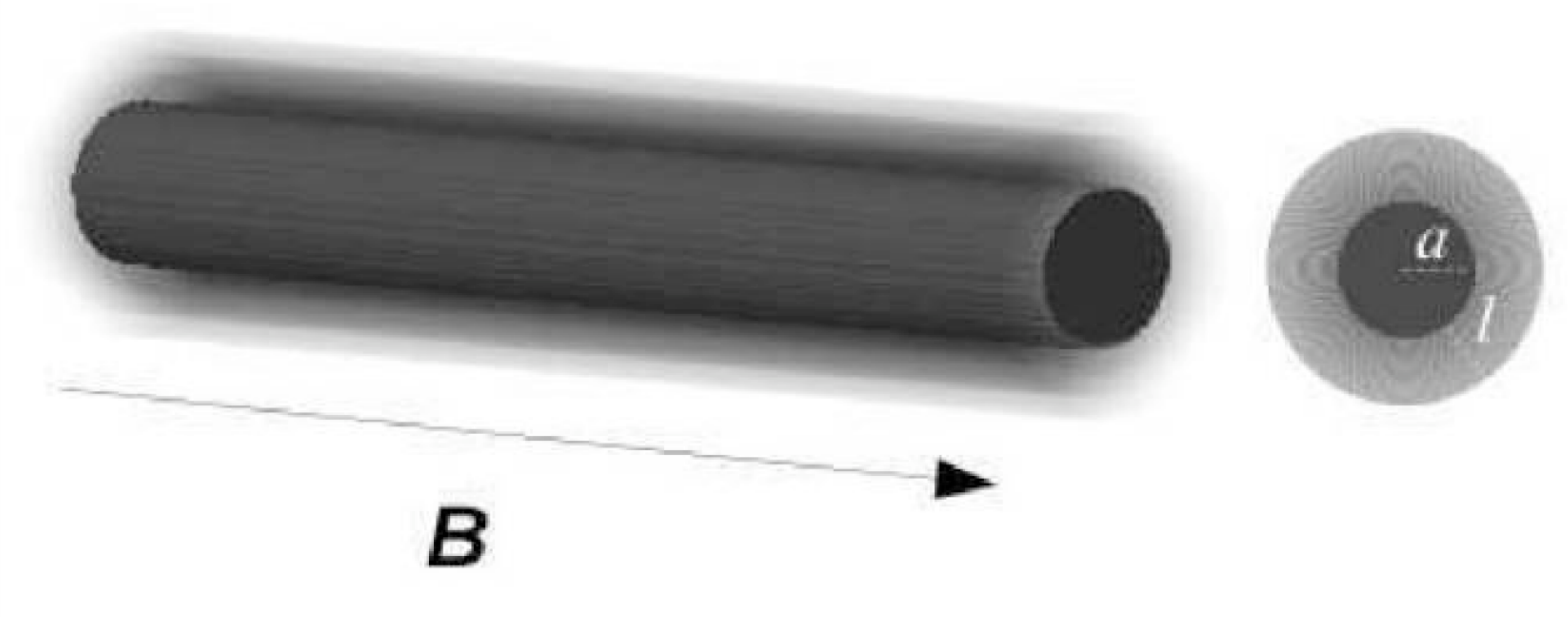}}
\caption{Sketch of a radially non-uniform filament fine structure of mean radius $a$, with a transverse inhomogeneous layer of thickness $l$. For $l=0$ the geometry of the model reduces to that of Figure~\ref{fig_nonadiab_thread}a. Figure taken from \cite{ATOB08}.}
\label{fig_thread_model}
\end{figure}

\begin{figure}[h]
\vspace{-6cm}\center{\mbox{
\includegraphics[height=10cm]{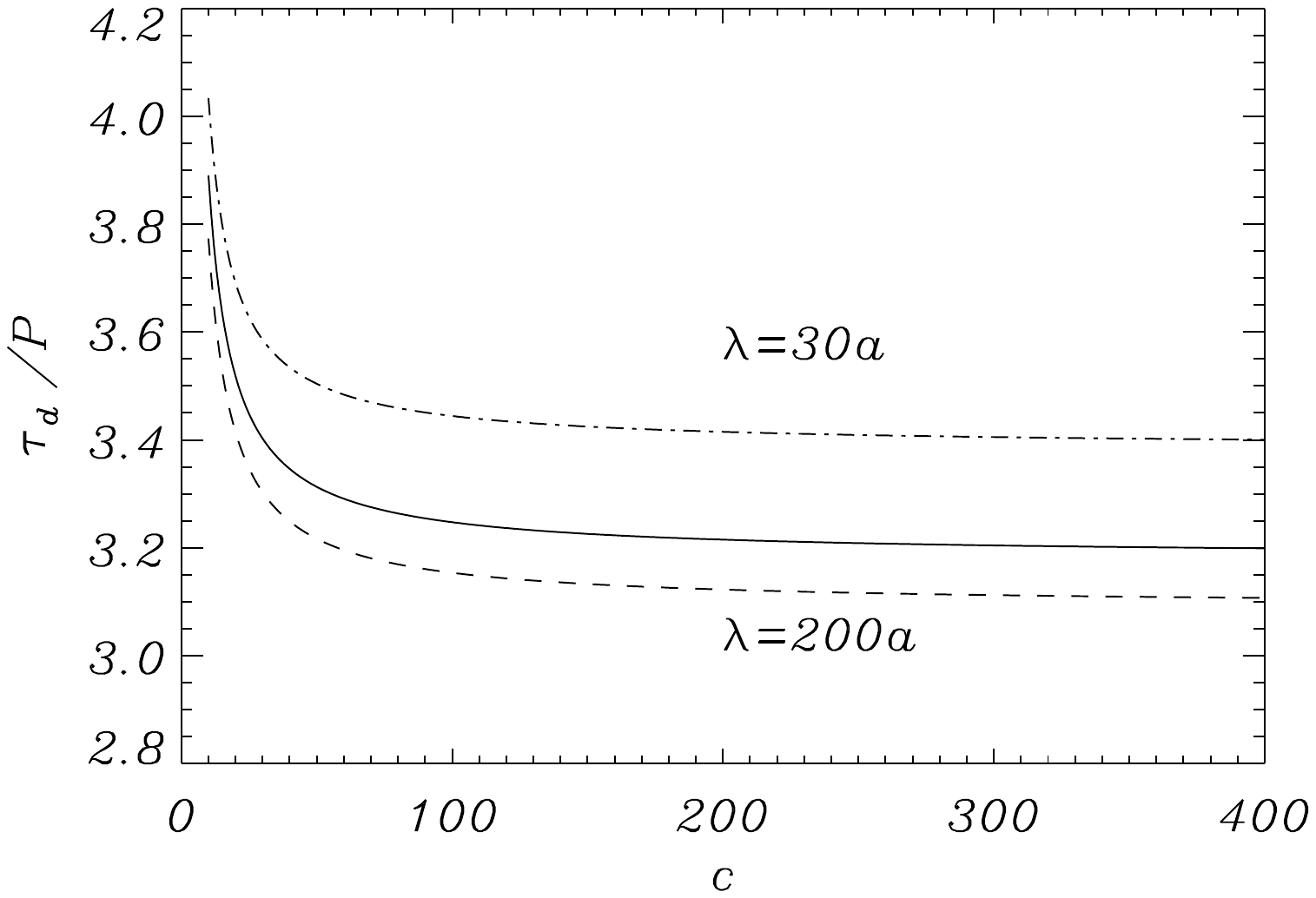}
\hspace{-2cm}\includegraphics[height=10cm]{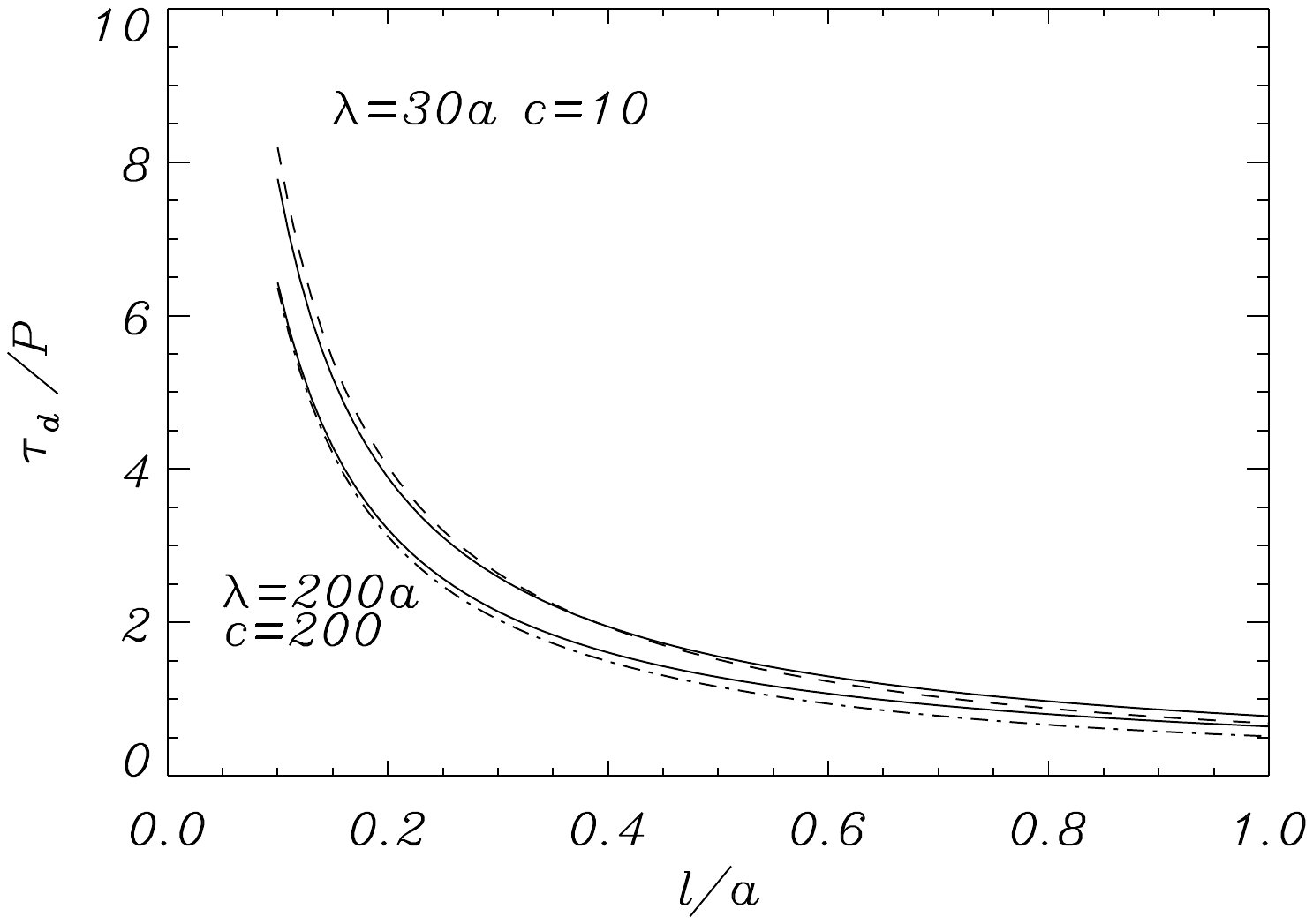}
}}
\caption{Ratio of the damping time to the period for fast kink waves in filament threads (radius = 100 km) attenuated by resonant absorption. (a) $\tau/P$ as a function of the density contrast with $l/a=0.2$ for two longitudinal wavelengths. (b) $\tau/P$ as a function of the transverse inhomogeneity length scale for two combinations of wavelength and density contrast. In both panels solid lines correspond to the analytical solution given by Equation~(\ref{fig_res_abs}) with $F=2/\pi$. Figures taken from \cite{ATOB08}.}
\label{fig_res_abs}
\end{figure}

The plasma density varies by about two orders of magnitude between a filament thread and the surrounding corona. In such a highly inhomogeneous configuration fast kink modes can be efficiently damped by transferring their energy to Alfv\'en modes through resonant absorption. \cite{ATOB08} examine the relevance of this mechanism in the attenuation of small amplitude prominence oscillations by modeling the magnetic and plasma configuration of an individual fibril as an infinitely long, straight, cylindrically symmetric flux tube of mean radius $a$ surrounded by a coaxial layer of thickness $l$ (Figure~\ref{fig_thread_model}). The $\beta=0$ approximation is used. The coronal and thread densities are denoted by $\rho_c$ and $\rho_f$, respectively, and their ratio as $c=\rho_f/\rho_c$. This parameter is much larger than unity and probably of the order of 100 or larger. In this simple model, the magnetic tube representing the thread only contains cool material, i.e. the presence of hot material inside the tube is ignored. Taking into account that the observed propagating waves in threads have a wavelength, $\lambda$, much larger than the tube radius, \cite{ATOB08} start by using the long-wavelength approximation to make some analytical progress. Next, they use the so-called thin boundary approximation ($l\ll a$) to derive the following expression for the ratio of the damping time to the period

\begin{equation}\label{eq_res_abs}
\frac{\tau}{P}=F\frac{a}{l}\frac{c+1}{c-1},
\end{equation}

\noindent where $F$ is a numerical factor that depends on the particular variation of the density in the nonuniform layer: for a linear variation $F=4/\pi^2$ \citep{HY88,GHS92}, while for a sinusoidal variation $F=2/\pi$ \citep{RR02}. A simple substitution in this formula of a density ratio $c=200$ and $l/a=0.1$, which corresponds to a transitional layer whose thickness is 10\% the thread radius, yields $\tau/P\sim 6$, in good agreement with observations. Then, this analytical approximation seems to indicate that the process of resonant absorption is an efficient damping mechanism for fast MHD oscillations propagating in these structures. Next, \cite{ATOB08} solve numerically the full set of linear, resistive, small-amplitude MHD wave equations for the transverse kink oscillations. In this manner, the approximations that lead to Equation~(\ref{eq_res_abs}) are avoided. These numerical calculations confirm the analytical estimate that, for the typical large filament to coronal density contrast, the mechanism produces rapid damping on timescales of the order of a few oscillatory periods only. They also reveal that for thin layers ($l/a\sim 0.1$) the inaccuracy of the long-wavelength approximation produces differences of up to $\sim 10$\% for the combination of short wavelength with high-contrast thread. For thick layers, differences of the order of 20\% are obtained \citep[in agreement with][]{vDAPG04}. Moreover, the authors conclude (as shown by Figure~\ref{fig_res_abs}) that the obtained damping rates are only slightly dependent on the wavelength of perturbations and that the damping rate becomes independent of the density contrast for large values of this parameter. This last result has two seismological consequences. First, the observational determination of the density contrast is less critical than it is in the low-contrast regime, the one corresponding to coronal loops. Second, according to seismic inversion results that combine the theoretical and observed periods and damping times \citep{AAvDGP07,GABW08}, high-density thread models would be compatible with relatively short transverse inhomogeneity length scales.

\begin{figure}[h]\sidecaption
\resizebox{0.5\hsize}{!}{\includegraphics*{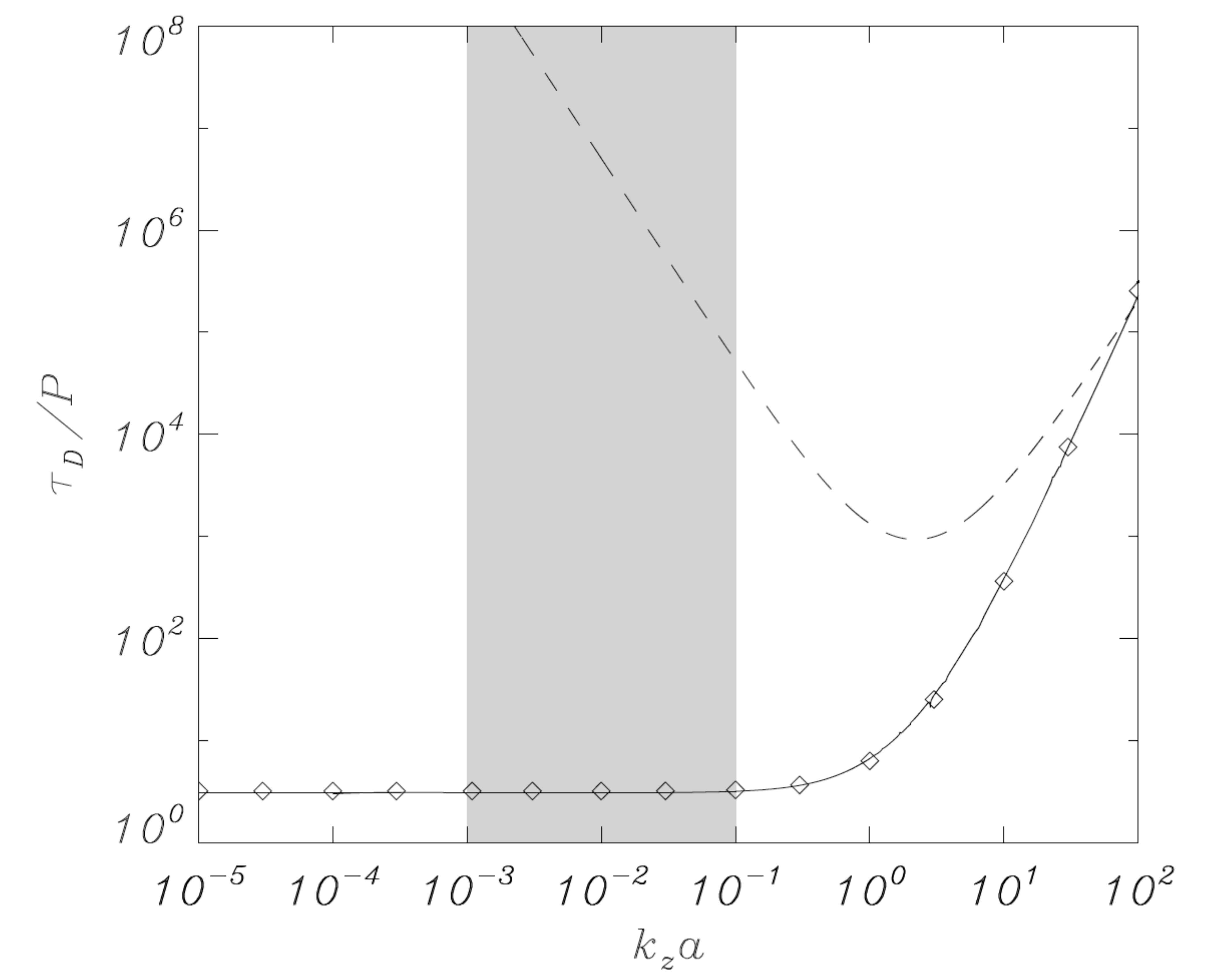}}
\caption{Ratio of the damping time to the period as a function of the dimensionless wavenumber, $k_za$, corresponding to the kink mode for $l/a = 0.2$. Attenuation in this cylindrical thread model is caused by resonant coupling to Alfv\'en and slow continuum modes. The solid line is the complete numerical solution, whereas the symbols and the dashed line are the results of the thin boundary approximation corresponding to the Alfv\'en and slow resonances, respectively. The shaded zone corresponds to the range of typically observed wavelengths in prominence oscillations. Figure taken from \cite{SOBG09}.}
\label{fig_slow_cont}
\end{figure}

In a subsequent work, \cite{SOBG09} note that in a $\beta\neq0$ plasma with prominence conditions embedded in the much hotter corona the fast kink mode can also undergo a resonant coupling to slow continuum modes. Hence, it is argued, the combined effect of the two resonant couplings may lead to an enhanced attenuation rate of the kink mode. The structure considered by \cite{SOBG09} is a thread identical to that of Figure~\ref{fig_thread_model}, although the $\beta=0$ assumption is removed from their calculations and, consequently, the plasma pressure is finite. Such as in \cite{ATOB08}, in this work a combination of analytical expressions and numerical analysis is used. The analytical formulas not only allow to predict the positions of the slow and Alfv\'en resonances, but they also provide the contribution of each of the two mechanisms to the damping time, $\tau_{DA}$ and $\tau_{Ds}$, whose ratio is

\begin{equation}
\frac{\tau_{DA}}{\tau_{Ds}}\sim\frac{(k_za)^2}{m^2}\left(\frac{c_s^2}{v_A^2}\right),
\end{equation}

\noindent with $m$ and $k_z$ the azimuthal and longitudinal wavenumbers and $c_s$ and $v_A$ the sound and Alfv\'en speeds. Using $m=1$ and a typical value for the wavelength ($k_za=10^{-2}$) results in $\tau_{DA}/\tau_{Ds}\sim10^{-7}$, which means that the resonant coupling of the kink mode with the Alfv\'en continuum produces an attenuation rate that is $10^7$ times smaller than that coming from the coupling with the slow continuum. These results are also verified from the numerical computations (see Figure~\ref{fig_slow_cont}), which confirm that resonant absorption with the slow continuum is irrelevant in the damping of the observed transverse thread oscillations.

\section{Conclusions}\label{conclusions}

Although the study of small-amplitude waves and oscillations in prominences constitutes a discipline that might complement the direct determination of prominence parameters by providing independent values based in the comparison between observations and theory, so far prominence seismology is more a promise than a reality because of the large gap between observation and theory. The analysis presented in Section~\ref{seismology} is a good example of a seismological application to prominences, but, unfortunately, it is the only one of its kind. Such a gap arises because of the little restrictions imposed by observational works (which sometimes reduce to reporting the period of the detected oscillations) and the simplicity of theoretical studies (which neglect the intricate nature of prominences and substitute it by a very simplified physical model). Nevertheless, these two sides are coming together as the complexity of works increases.

In this paper the possible role of several physical mechanisms in the attenuation of prominence oscillations has been reviewed. It has been shown that both the fast and slow waves can undergo strong damping under a variety of physical conditions and with the intervention of different effects. Nevertheless, the list of the presumably relevant mechanisms has not been totally explored and other effects should be investigated. In addition, the geometry of the models is too simplified, which implies that the results derived so far only give a rough guide to the damping features in actual prominences. The efficiency of wave leakage and coronal conduction in draining oscillatory energy from the prominence and transferring it to the corona calls for the study of complex configurations in which the structure of the full prominence-corona system is treated more realistically.

Most of the works are based on the linear approximation, whose validity seems quite robust in view of the small amplitude of prominence oscillations in many cases. Nevertheless, apart from the so-called small-amplitude prominence oscillations, that in general affect small regions of a prominence and have Doppler velocity peaks typically below 1--2 km s$^{-1}$, there is a different kind of phenomenon, the large-amplitude oscillations (Tripathi et al. 2009, this volume). They are characterised by a much higher amplitude (with oscillatory velocities up to 90 km~s$^{-1}$) and disturb the whole filament; see \cite{eto_etal02,jing_etal03,okamoto_etal04,JLSW06,IT06,VVTZ07}. These prominence vibrations also damp very rapidly, in a few periods, so it is in order to ascertain whether the mechanisms at work in the damping of small-amplitude oscillations are also the main ones for the large-amplitude counterparts or whether non-linear effects can also cause the attenuation of the latter.

\begin{acknowledgements}
The author acknowledges the International Space Science Institute for the financial support received to attend a meeting of the team ``Coronal Waves and Oscillations'' in Bern. Financial support received from the Spanish MICINN and
FEDER funds, under the grant MEC AYA2006-07637, and from the Conselleria d'Economia, Hisenda i Innovaci\'o of the Government of the Balearic Islands, under grant PCTIB-2005-GC3-03, is also acknowledged.
\end{acknowledgements}


\end{document}